# Life's Solutions are Complex Fluids

## *A Mathematical Challenge*

Bob Eisenberg
Department of Molecular Biophysics and Physiology
Rush University and Medical Center
Chicago IL
*beisenbe@rush.edu*

August 30, 2012

*Typos corrected, I hope.*

## Abstract

Ionic solutions are so important that they are often ignored. Chemical reactions have been studied for more than a century in ionic solutions. Life occurs in electrolyte solutions made of mixtures of 'bio-ions' (sodium $Na^+$, potassium $K^+$, calcium $Ca^{2+}$, and chloride $Cl^-$), along with many other charged components. The precise composition of biological solutions is important. Gradients of $Na^+$, $K^+$, $Ca^{2+}$, and $Cl^-$ ions provide energy to drive signals through the nervous system, and energize transport in most cells. The signals of biology ('biosignals') themselves are usually ions, often just $Ca^{2+}$. Ionic concentrations range from $10^{-11}$ M to $10^{-1}$ M to $10^2$ M in different biological systems, from hormones, to Ringer solutions, to solutions in and near enzyme active sites, binding proteins and ion channels. The electric field of ions is always important. Ions interact with each other, and everything else that is charged, to maintain global electroneutrality. Otherwise, electrons are stripped off atoms and bio-compounds are destroyed. Bio-solutions contain a myriad of other charged compounds from simple organics like carbonate, to organics like ATP, and macromolecules like DNA, RNA, and proteins; even the organelles of cytoplasm are themselves charged, rather like colloids. Nearly all components of living solutions are charged. All interact through the electric field. Everything interacts with everything else.

Classical thermodynamics and statistical mechanics describe systems in which nothing interacts with nothing. Even the highly refined theory of simple fluids does not deal very well with electrical interactions, boundary conditions, or flows, if at all. Electrical interactions, boundary conditions, and flows are essential features of living systems. Life without flow is death and so a different approach is needed to study biology alive.

The theory of complex fluids deals with interactions, boundary conditions, and flows quite well as can be seen in its successful treatment of liquid crystals.

I advocate treating ionic solutions in general as complex fluids, with microelements that are the solutes and components of the solution. Enzyme active sites are a special case where some solutes are reactants. Solutes are crowded into active sites of enzyme by the high density of protein charges. The electric field links chemical reactions to charges in the protein and surrounding solutions. Interactions potentiate catalysis and control biological function.

I suspect that most chemical reactions that occur in liquids also need to be treated by the theory of complex fluids. The electron movements of these reactions occur in a temporary highly concentrated fluctuation, a transient spatial inhomogeneity in the bulk solution. The electron movements of these reactions (described by quantum mechanics) are coupled to the electric (and sometimes steric) fields of the bulk solution. I suspect the electron movements, inhomogeneities, and chemical reaction (in the condensed phase) need to be treated by the theory of complex fluids because everything interacts with everything else, in this system, as in so many others.

**It is hard to see big things from up close**, in math and science, as in the world. Many mathematicians approach biology for interesting problems nowadays but sometimes the biggest problem, with the greatest potential, is too close to see.

All of biology occurs in salt solutions evolved from the primitive oceans of the earth [241, 240, 139, 473, 474]. Water without the ions of the ocean is lethal. Almost all cells burst when exposed to pure water. Most enzymes denature in distilled water. The need for ions in water is thus general in biological systems. The need for ions is specific (as well as general). Not just any ion will do. Most biological systems require quite specific salt solutions to function and those salt solutions are almost always mixtures involving definite concentrations of $Na^+$, $K^+$, $Ca^{2+}$, and $Cl^-$ ions.

**Ions in water are life's plasma** [148]. The ions in biological mixtures (called 'Ringer solutions' in general) carry information that controls biological systems. The selective flow of some of these ions are the signals of the nervous system. The selective flows of ions coordinate contraction of muscle. They allow the heart to function as a pump. The concentration of ions is important. Concentration often regulates function the way the gas pedal in a car controls its speed. The location of ions is important. The same ion (often divalent calcium $Ca^{2+}$) in different locations is a different signal, even if both locations are in the same cell. The biological role of ions is a central subject in physiology and medicine [61].

**Selectivity.** Different ions have different specific biological functions just as signals in different wires in a computer have different specific functions. Life cannot be understood without knowledge of the diversity of chemical signals, just as a computer cannot be understood without knowledge of the voltages in its wires. These signals are a central subject in biochemistry, cell and molecular biology, and medicine [7, 470, 492]. The diversity of chemical signals depends on the selectivity of biological systems to ions. Different ions are recognized selectively by the proteins of life, whether those proteins are ion channels or enzymes.

The origin of selectivity has been considered mysterious. Indeed, the origin of specificity has been called the central question of life, by Nobel prize winner Aaron Klug [402]. The origin of selectivity in many systems is not known, or is in dispute, but in some cases selectivity can be calculated and results of experiments predicted before they are performed, treating ions with physical theories developed to deal with mixtures of ions in non-biological systems. There are three ion channel proteins (of great importance throughout biology) where this approach has been successful, over a wide range of conditions [372, 201, 497, 373, 496, 517, 58, 57, 191, 193, 194, 196, 59, 154, 197, 314, 432, 55, 141, 221, 313, 121, 195]. The inverse problem has been solved showing how to determine the internal charge structure of a channel protein from measurements of current voltage relations, in the presence of random and systematic error [18, 19, 66] .

**Ionic mixtures** have been studied experimentally in some detail since around 1900. Kraus [312] provides a fine summary of the classical literature written for mathematicians. Laidler [321] et al provides a good introduction to physical chemistry for mathematicians. Fawcett [165] is a useful clear textbook of electrochemistry. The first pages of Fraenkel [173] summarize the present state of knowledge, as I see it too.

After all this time, one would imagine that the specific properties of ionic mixtures of such significance to biology would be understood. But they are not. This fact is hard to believe, particularly for workers in adjacent sciences, where complex phenomena of heterogeneous fluids are computed with striking success by computational fluid dynamics [11, 522, 528], able to compute very complex phenomena indeed, from ocean waves [339] to liquid crystals [190, 132, 133]. One of the motivations for this paper is simply to document the lack of understanding of life's solutions, of ionic mixtures, as stated so clearly in the literature of physical chemistry.

**The lack of success** in computing the most elementary properties of ionic solutions is in striking contrast to the range of accurate experimental data available for many years, reaching back to the 1930's, and even earlier [312, 235, 420, 176, 31, 416, 29, 410, 175, 477, 174, 259, 399, 468, 520, 324, 406, 316, 69, 398, 93, 407, 137, 348, 33, 136, 317, 321, 165, 424, 233, 285, 298, 299, 297, 403, 224, 307, 309, 56, 74, 286, 323, 3, 124, 239, 270, 277, 308, 318, 319, 335, 336, 356, 495, 151, 173, 172, 262, 295, 296, 436, 519, 150, 141, 189, 264, 378, 413, 412, 523, 191, 198, 199, 200, 491, 232, 391, 417, 421, 437, 122, 135, 462, 41]. The gap between data and theory provides an important opportunity for the present generation of scientists, particularly given the biological and technological significance of ionic solutions. I believe that advances in mathematics and computational science warrant a fresh look at ionic questions.

**The question is where to start?** when dealing with all this experimental data.

The place to start is calculations of current voltage curves in standard electrochemical cells described in textbooks of electrochemistry [28, 52, 53, 165, 321, 391, 430, 441, 442, 516] and polarography [63, 414]. Current voltage relations are the fundamental data from which most of the classical properties of ions (used to describe biological systems) are derived. The most important derived quantity is the activity of ions, the generalization of number density or concentration appropriate for nonideal solutions. The activity of ions plays the role of height and weight in the gravitational field. It is the fundamental determinant of energy. Gradients of activity drive flow. The activity equals the concentration in infinitely dilute solutions of noninteracting particles called ideal solutions or perfect gases.

**Ions are a compressible plasma** within the incompressible liquid formed by water and ions [155]. The concentration and type of ions varies enormously in biology. Evidently, evolution has used this variation to produce and control most biological functions. Electrodiffusion is to life what current flow is to computers. It is the fundamental process in nearly everything the system does.

The free energy per mole, the activity of the ions, can serve many biological functions. In trace concentrations (smaller than say $10^{-7}$ M), concentration of a particular ion (in a particular place in a particular biological cell, or subcellular organelle) is a specific signal. In large concentrations (e.g., the 0.2 M solutions of sodium ions $Na^+$ outside cells), the ions serve as sources of (free) energy for many processes of great importance, like signaling in the nervous system. In enormous concentration (say 20 M) ions create a bizarrely specialized environment (in ion channels [141] and active sites of enzymes [159, 280]), more like an ionic liquid [13, 37, 309, 404, 438, 514] than an ideal solution. In channels, this environment allows specificity and control in a protein nanovalve. In enzymes, this environment creates an electrostatic glove wrapping the chemical reactants and must have a decisive importance in the mechanism of catalysis [511, 512, 459, 509, 502], in my opinion. I hasten to add that how this electrostatic glove is important in catalysis remains a mystery, at least to me. Channels and enzymes are intimately related [159, 147], variations on a theme, intertwined as obviously and as mysteriously and as successfully as themes in Bach's F major Toccata (BWV 540).

The free energy per mole of ionic solutions is difficult to calculate. It is difficult to calculate the properties of pure solutions of monovalents (e.g., $Na^+Cl^-$) over a wide range of concentrations, particularly at high concentrations. Calculations of pure divalents (like $Ca^{2+}Cl_2^-$) are even more problematic. The activity of ions in mixtures particularly as they flow cannot be described at all well by existing theories. References supporting this strong statement include [520, 495, 477, 468, 410, 408, 407, 398, 399, 335, 321, 319, 318, 316, 312, 296, 286, 284, 270, 262, 259, 235, 175, 176, 172, 165, 137, 136, 124, 42, 41, 40, 38, 37, 33, 4, 3].

**Ionic solutions are not ideal.** The behavior of ions is nothing like the behavior of infinitely dilute ideal solutions (of uncharged noninteracting particles) assumed in textbooks of biochemistry and physiology, even in elementary textbooks of chemistry and electrochemistry. The behavior of ions is better captured by chemical engineers [308, 520], who need to compute the properties of electrolytes and mixture if they are to deal with salt solutions in their flow reactors, chemical plants and factories. The empirical formulations used by chemical engineers, however, are not derived from physical models. They include many parameters that cannot be transferred from one condition to another. The parameters need to be modified when the models are applied to new conditions [276, 343, 384, 452]. How to modify the parameters is not known. So I conclude the theories of chemical engineers do not work very well.

Only a few of the empirical formulations of chemical engineering apply to flow [287, 421]. The properties of ionic solutions most important for life occur in flowing solutions. These properties are not addressed by classical thermodynamics or statistical mechanics, or even by the theory of simple fluids, for the most part. Not addressed, the problems posed by flow cannot be solved. In biology, flow ceases only with death. Classical methods treat ionic solutions as very complicated simple fluids. Classical methods are not powerful enough to deal with flow.

I am aware of the difficulties and challenges involved in attacking this long lasting problem. After decades (approaching a century) of efforts by many of the most able physicists, including Lorentz, Debye, Onsager, Kirkwood, and so on, there is no satisfactory theory for salt water or the closely related ionic mixtures inside animals and plants even when flows are identically zero, at thermodynamic equilibrium. There are essentially no theories—satisfactory or not—that deal simultaneously and self-consistently with the convection, diffusion, and migration, along with volume regulation of cells, vital to the function of kidney, heart, lungs, etc. Such models must include single atom ions (like sodium $Na^+$, potassium $K^+$, calcium $Ca^{2+}$, and chloride $Cl^-$ that I call '*bio-ions*' because of their biological importance), complex organic molecules that are ions like ATP, sugar acids, carboxylic acids, and bioamines, all of which have essential roles, along with nearly all of the organic chemicals and polymers described in biochemistry texts. They are all ions, of some complexity, most with complex structure and internal dynamics obviously involved in their biological function. The tools of classical physical chemistry based on ideal solutions at thermodynamic equilibrium are unequal to the task of describing let alone understanding and predicting the properties of these solutions, in my opinion.

**Understanding the physical properties of these ions forms a central problem** in all of biology, unsolved, in my view, because of the lack of tools.

Modern analytical methods (of the self-consistent theory of complex fluids, for example) and modern numerical and computational methods (made possible fundamentally by 50 years of diligent exploitation of Moore's law [351, 380, 381]) can attack and probably solve these problems. But first the mathematicians who know these methods must learn of the problems.

If mathematicians analyze the wrong equations, their results will be less helpful than they wish, not worthy of the efforts involved.

Mathematicians are understandably more interested in the solution of equations than in the solutions of life. Properties of ionic solutions—often messy—are less interesting to them than properties of mathematical solutions—ofen elegant. Thus, it is entirely understandable that mathematicians have concentrated almost all their efforts on the Poisson Boltzmann PNP (Poisson Nernst Planck) system. After all the PNP system has been remarkably successful in dealing with semiconductor systems [486, 357, 228, 358, 451, 363, 279, 242, 352, 263, 488] and it is natural to hope that success would extend to ions in solutions and biological systems [127, 129, 128, 331, 330, 328, 329, 358, 327, 65, 107, 334, 332, 109, 108, 430, 326, 325, 25, 27, 81, 82, 157, 39, 75, 160, 158, 26, 78, 79, 83, 96, 146, 145, 338, 395, 76, 95, 117, 131, 161, 192, 252, 320, 333, 394, 8, 70, 112, 118, 152, 223, 253, 275, 383, 94, 97, 111, 272, 396, 448, 481, 98, 99, 115, 140, 198, 251, 254, 273, 274, 449, 482, 483, 116, 153, 144, 360, 385, 484, 5, 38, 6, 101, 113, 119, 149, 349, 494, 114, 361, 364, 439, 493, 19, 66, 100, 156, 2, 142, 143, 463, 18, 36, 350, 464, 347, 141, 337, 523, 524, 121, 475]. Much other literature is cited in [38, 160, 158].

**PNP treats ions as points.** The sad fact is, however, that PNP treats ions as points, and that treatment is inadequate for ion solutions. The finite diameter of ions makes the electric field very different from that of points. The crowding of ions makes the entropy very different. And the fact that different ions have very different diameters means that complex layering phenomena characterize mixtures of ionic solutions. No single 'distance of closest approach' can do. Different ions have different diameters. So different ions have different distances of closest approach. Layers of different charge density and sign can and do result.

Equations that deal with ions as points miss these phenomena. Biology and evolution have not missed these phenomena. Biology and evolution use the finite diameter of ions to build systems that allow animals to survive. And equations must deal with the finite diameter of ions if they are to allow biologists to understand life.

**Biological systems cannot be understood if ions are treated as points.** PNP equations will miss the problems that matter most to living systems because the PNP equations leave out most of the nonideal properties of ionic mixtures, particularly those containing divalents, like seawater and the solutions inside animals.

I believe that an ionic solution should be viewed as a specific type of complex fluid that couples hydrodynamics to electrostatics and to the microstructure of charged particles, their excluded volume and even their shape. Ionic solutions are complex fluids in which atoms interact with nearby confining structures through several types of forces. Ionic solutions are complex fluids in which the behavior of individual atoms and proteins (e.g., ion channels) is changed by charge on far distant boundaries. I believe that existing methods of the self-consistent theory of complex fluids will allow rapid progress on previously intractable problems.

The mathematical treatment of ionic mixtures I think appropriate is that of complex fluids and its close cousin the theory of transport in semiconductors. As many mathematicians know very well, the theory of complex fluids is designed to deal with fluids with interacting microelements that involve many types of physical forces and fields, that interact across all scales [258, 265, 341], ranging from atomic scale issues of excluded volume to the macroscale fields of electrostatics, even involving boundary conditions 'at infinity' that can couple to the atomic scale, as they do to create the propagating electrical signals of the nervous system.

**'Plasmas of life' and the solutions of chemistry** can be viewed productively as complex fluids, in my view. The bio-ions of biological plasmas themselves are (relatively simple) microelements that perturb the electric field by their finite size, and introduce other steric constraints. The long range of the electric field guarantees that 'everything interacts with everything else'. The significant size of the ions distorts the electric field substantially. The impenetrability of the ions (see p. 630 of [44]) implies steric constraints that dramatically change the entropy and free energy of concentrations of monovalents greater than $\sim 100$ mM, or mixtures, particularly those involving divalent ions. The theory of simple fluids is not able to deal well with these complexities as they are found in experiments. A fully self-consistent treat of complex fluids can only do better, I think.

**Molecules in solution in general are not as simple as bio-ions.** Molecules are microelements that have complex shape and internal dynamics. Molecules interact with all other elements and fields on several scales. Polymers, including proteins, are macromolecules that involve atomic scales and also macroscales. Side chains of the polymer can be micro̲elements themselves. Think of side chains of proteins. One polymer molecule itself can also be a macro̲element cm in size (think of our finger nails).

Side chains of proteins are often acids (i.e., have permanent negative charge) or bases (i.e., have permanent positive charge) that mix in an electric stew [365] with water and bio-ions. Polymers are both micro and macro-elements that interact dramatically with other ions through the electric field and steric constraints, as well as their internal properties. Indeed, the organelles of cells (mitochondria, ribosomes, nucleoli, etc.) are macroelements that form crucial components of the complex fluid inside cells that biologists call cytoplasm.

Outside cells, bio-ions are not alone. In blood, bio-ions are part of what biologists call plasma (when mixed with glucose). Biological plasmas form the highly complex fluid of blood when they mix with a zoo of organic molecules and cells like red and white blood cells and many others. If anything needs analysis as a complex fluid, it is surely blood!

**Chemistry occurs in complex fluids.** One can argue that classical chemistry (that occurs in the liquid phase) occurs in a complex fluid and needs to be studied as liquid crystals are, by the theory of complex fluids.

Chemistry is about chemical reactions, in which electrons change their relation to nuclei, and so change the properties of molecules dramatically. Chemists have focused their attention on the molecule, atom, and electron for some one hundred and fifty years, so they could develop chemical knowledge and technology. Chemistry textbooks usually ignore the fluid in which chemical reactions occur. This was just as well since mathematical methods to deal with interactions of chemicals and surrounding fluids were not known until very recently.

**Chemical reactants in general can be viewed as microelements in a complex fluid.** The theory of complex fluids guarantees that models involving chemical reactions and the fluid will be self-consistent. Models of each microelement (reactant) will be needed but models involving all microdynamics of reactants are likely to be intractable. But reduced models of these reacting molecules are often already known. Those have been the output of research in the chemical sciences for a very long time. These models of reactants can be formalized and perhaps improved by appropriate extensions of the theory of inverse problems, I suspect. The theory of inverse problems can help identify the variables that determine macroscopic flows and separate them from the variables that do not have large effects. Reduced models are models that account for the important variables that determine macroscopic flows and phenomena.

Chemical reactions can be treated by the theory of complex fluids applied to reduced models. I imagine the theory will include interactions of microelements (reactants) involving rearrangements of internal (electronic) structures of reactants, according to Schrödinger's wave equation of the electron, coupled to the local transient highly concentrated environment of ions and water in which the reaction occurs, and to the ions and water (and boundary conditions) of the bulk solution in general.

**Dilute ionic solutions.** Fortunately, the study of bio-ions does not depend on the study of the Schrödinger equation. Bio-ions can be studied by classical methods of physical chemistry, starting with the theory of dilute ionic solutions.

Historically, the mathematics of ***dilute*** ionic solutions is often idealized by the family of Poisson Boltzmann equations, starting (as far as I know) with the Gouy Chapman and Debye Hückel theories of a century ago, more or less. (Dilute ionic solutions have concentration below 1 mM, if one adopts the stringent view of experimental physical chemists, p. 55 of [520]. A less stringent concentration limit is good enough from my point of view, say 50 or even 100 mM.)

Reincarnated as the Poisson Boltzmann or (in different form) as the Born model of ionic solutions, such equations have also received a great deal of attention as beginning models of proteins in biological systems [505, 502, 498, 500, 509, 510, 210, 217, 411, 431, 213, 303, 212, 236, 64, 208, 214, 218, 513, 209, 257, 506, 126, 397, 455, 454, 211, 504, 508, 453, 256, 16, 219, 14, 17, 204, 237, 255, 15, 515, 206, 207, 171, 222, 501, 507, 134, 188, 355, 479, 489, 88, 90, 125, 203, 354, 362, 428, 450, 490, 89, 202, 289, 353, 460, 503, 24, 67, 73, 130, 205, 271, 392, 499, 72, 84, 87, 167, 292, 374, 379, 459, 281, 471, 476, 527, 85, 138, 288, 290, 512, 71, 86, 220, 302, 346, 415, 511, 10, 245, 282, 291, 301, 335, 336, 375, 467, 525, 91, 173, 172, 215, 296, 344, 390, 92, 216, 376, 466].

**Poisson Boltzmann equation**. The mathematical properties of the Poisson Boltzmann equation have been reviewed recently in SIAM Review [518], which can serve as an entry to this immense literature, as can [38, 160, 150]. It is gratifying to see that the power of modern computational mathematics is being focused on ionic solutions [518, 335, 336, 523, 92, 524, 68] and much more work no doubt that I do not know about. This is certainly an essential first step in applying mathematics to the role of ions in biology.

But the Poisson Boltzmann models—even analyzed with powerful computational mathematics—are only a beginning. These models have a severe limitation. They treat ions as points. The Poisson Boltzmann family of equations is too crude to deal with the concentrations of monovalent ions like sodium, potassium and chloride that occur in biology. These equations fail altogether for the divalent ions like calcium ions that play such an important role throughout biology. These equations do very poorly for the mixtures that are the plasma of life (and the seawater of our oceans). In fact, it is well known, as we have already said, that Poisson Boltzmann is only valid for solutions of one type of monovalent (e.g., sodium chloride), in concentrations below say $10^{-1}$ M [312, 235, 420, 176, 31, 416, 29, 410, 175, 477, 174, 259, 399, 468, 520, 324, 406, 316, 69, 398, 93, 407, 137, 348, 33, 136, 317, 321, 165, 424, 233, 285, 298, 299, 297, 403, 224, 307, 309, 56, 74, 286, 323, 3, 124, 239, 270, 277, 308, 318, 319, 335, 336, 356, 495, 151, 173, 172, 262, 295, 296, 436, 519, 150, 141, 189, 264, 378, 413, 412, 523, 191, 198, 199, 200, 491, 232, 391, 417, 421, 437, 122, 135, 462]. And of course the Poisson Boltzmann treatments do not deal with flow at all, not even with the stationary (tracer) unidirectional fluxes that define active and passive transport in biological systems [247, 249, 250, 22, 23, 48, 61] as reviewed in the useful historical collection [478] and analyzed by mathematician Ludwig Bass [34, 35, 368, 369].

The powerful mathematics reviewed in [518] needs to be applied to more realistic models of ionic solutions to be useful in dealing with biological function in general. Specific experimental situations certainly exist in which the Poisson Boltzmann approach is useful, but natural biological function almost always occurs in solutions beyond the reach of the Poisson Boltzmann family of equations.

**Limitations of the Poisson Boltzmann** family of equations are well known to physical chemists and are identified and discussed in innumerable references, including [312, 235, 420, 176, 31, 416, 29, 410, 175, 477, 174, 259, 399, 468, 520, 324, 406, 316, 69, 398, 93, 407, 137, 348, 33, 136, 317, 321, 165, 424, 233, 285, 298, 299, 297, 403, 224, 307, 309, 56, 74, 286, 323, 3, 124,

239, 270, 277, 308, 318, 319, 335, 336, 356, 495, 151, 173, 172, 262, 295, 296, 436, 519, 150, 141, 189, 264, 378, 413, 412, 523, 191, 198, 199, 200, 491, 232, 391, 417, 421, 437, 122, 135, 462, 41]. As recently stated by leading experimentalists,

> **"It is still a fact that over the last decades, it was easier to fly to the moon than to describe the free energy of even the simplest salt solutions beyond a concentration of 0.1 M or so."**

This quotation states the common knowledge of the physical chemistry community for many decades, earlier stated, for example, by Torrie and Valleau [477], referring to the Poisson Boltzmann family of models:

> **"It is immediately apparent that classical theory has broken down completely. It …. fails to show [the] qualitative behavior [and] is seriously in error for quite low concentrations and charges".**

In verbal discussions, physical chemists customarily describe theories of point particles with language that reflects their feelings more than their rational analysis. Theories of ionic solutions that describe ions as points elicit an emotional response from physical chemists, because such theories have been known to be inadequate for nearly a century. The fundamental difficulty is that classical theories are obviously unable to fit data at any useful concentration, see the text of Barthel. *et al.,* [33] who say on p. 325, with slight paraphrase,

> **"Theories with point ions are restricted to such low concentrations that their experimental verification often proves to be an unsolvable task."**

The classical text of Robinson and Stokes ([420] not otherwise noted for its emotional content) gives a glimpse of these feelings when it says on p. 302

> **"In regard to concentrated solutions, many workers adopt a counsel of despair, confining their interest to concentrations below about 0.02 M, … "**

Theories fail altogether in the view of these experimentalists [33], [420] unless they deal with screening and also include a distance of closest approach.

The despair comes (I imagine) from the immediate realization that a distance of closest cannot be defined uniquely when different types of ions of different diameter are present, and so theories that describe ions as points are either unable to deal with data (if they do not include a distance of closest approach) or illogical (if they try to describe mixtures of bio-ions, like biological solutions, with a single distance of closest approach).

Feelings are made worse when one realizes that all biology occurs in mixtures of ions of quite different diameter. Ions of quite different diameter are likely to produce layering of charge in their ionic atmosphere. Charge is likely to layer in concentric shells around a central ion, if the ions have different signs of charge, or even just different charges (like $K^+$ and $Ca^{2+}$). In these cases, the layering will produce complex electric fields, allowing rectification, and other phenomena of great complexity , as layering does near charged walls [267, 31, 37, 260, 262] and in semiconductors. Layering of charge is responsible for most of the remarkably useful nonlinear properties of semiconductor devices, including transistors and the full range of nonlinear devices that make our digital technology possible [457, 472, 363, 488, 147]. The frustration of dealing with an illogical inappropriate theory that obscures such important phenomena tends to produce an emotional response among physical chemists, as it would in most of us.

Many review papers and textbooks have quotations of similar pungency (e.g., [33, 165, 173, 321, 324, 323, 468]). Many physical chemists have used far stronger language in private than I think appropriate to reproduce here. They particularly are concerned by the motivation of workers in other fields who ignore nearly one hundred years of experimental physical chemistry, all of which comes to the same conclusions, just discussed, and is easily accessible in papers, reviews, monographs, and textbooks.

It is understandable that mathematicians are unfamiliar with the reality of ionic solutions, despite the physical chemists' knowledge and feelings. The problems with point theories arise when fitting real data, which is complex and not at all ideal. The experimental setups that measure nonideal properties of ions are intricate. The variables used to describe the ionic solutions are not familiar and the literature is hard to deal with in practical library terms. It is very large (see compendia [235, 420, 106, 410, 259, 520, 398, 308, 318, 264]) but it is hard to find in computer archives because so many of the measurements were made before 2000.

**Mixtures like biological solutions are never ideal or simple.** The experimental work shows very clearly, however, that nonideal properties are particularly important in mixtures like sea water and the various 'Ringer' solutions outside and inside cells, where calcium ions always play an important role, and in the highly concentrated solutions in and near DNA, enzyme active sites, ion channels, and the electrodes of electrochemical systems. The first pages of Fraenkel [173] states the situation well in a few paragraphs.

The most important reason for the failure of the Poisson Boltzmann theories is simple. Those theories treat ions as points, but in almost all solutions the size and shape of particles are important. The chemical tradition is based on the theory of simple fluids built on the understanding of ideal gases [60, 429] which are infinitely dilute without interactions. The modern theory forms a beautiful highly refined body of work [29, 32, 234, 233, 417], but it still does not easily accommodate long range fields, and their boundary conditions, finite size (because the closure problem remains unsolved), or multiple body interactions. Nonuniform (spatial) boundary conditions, and flow are not present in most of these theories despite their importance in technology and biology. Even extracellular solutions in the biological context are too complex to be viewed as simple fluids, in my opinion, because they are (nearly) always away from equilibrium, with flows of some component or other, and are mixtures involving long range forces, boundary conditions, and multiple body interactions of impenetrable bio-ions.

**Not all biological solutions are as simple as extracellular solutions of bio-ions.** Not all extracellular solutions are so simple. Most biological solutions contain organic molecules made of many atoms joined by covalent bonds and so have complex shapes and exceedingly complex movements and internal 'vibrations' as well. These molecules are so complex and interesting from a chemical point of view that the entire profession of organic chemistry has been devoted to their study for nearly 150 years. (A glance at the literature of organic chemistry, through a search of the internet, shows a depth and complexity approaching that of the entire discipline of mathematics!) In the simplest context, think of ATP, or amino acids, or even the natural bicarbonate buffer found throughout life (unavoidable because the $CO_2$ of the atmosphere dissolves into any solution exposed to air). All of these molecules consist of many atoms joined by covalent bonds with various properties that allow a wide variety of internal motions, both vibrations and rotations. All of these molecules have many atoms with a definite permanent charge, so the atoms interact by the electric field directly, by the properties of their covalent bonds, and by their resulting impenetrability. Any solution of an organic molecule is a highly complex fluid.

**Internal dynamics of organic molecules are important**, as they are in the microelements of many complex fluids, if the molecule is more complex than a sphere. Many of the important molecules dissolved in a biological plasma are much more complex than a sphere. All the metabolites of classical biochemistry are organic molecules of some complexity. Nucleic acids and proteins are polymers with internal motions of great complexity. Clearly, the internal dynamics of metabolites and polymers need to be included in a complete description of a biological plasma. No one has known how to do that (historically: [139, 473, 474]). Treating these systems as microelements in a complex fluid represents a new departure of great potential.

**Even the simplest solutions need a theory of complex fluids.** The electric field of the ionic spheres depends on their diameter; the steric exclusion of the spheres involves large excess (free) energies because the number density of these spheres is very large where it matters. Even in solutions like seawater, the number density in the ionic atmosphere *near each ion* is large enough to make the finite size of ions important. The electric field near a dense mixture of spheres is quite different from the electric field of an infinitely dilute set of points. The steric repulsion of the crowded sphere is also important. The inner shell of the ionic atmosphere would be very different if ions were points.

**Free energy of one type of ion depends on the number density of _ALL_ other types of ions** in real ionic solutions. This is the fundamental fact that makes simple treatments of ionic mixtures so difficult. The idea that ions move independently in bulk solution is simply false. Only under the most extreme conditions (e.g., monovalents below 1 micromolar concentration) is independence a decent approximation. Even at 1 millimolar concentration the ions of a monovalent salt like $Na^+Cl^-$ are strongly coupled by the electric field and do not behave independently at all. That is exactly why PNP and Poisson Boltzmann were introduced.

In biological conditions, ionic solutions are nothing like ideal. Everything interacts with everything else. The 'driving force' for the movement of any type of ion depends on the concentration (and perhaps flow) of all other types of ions.

This is an experimental fact apparent in the tables of measurements from innumerable laboratories [41, 106, 235, 259, 264, 308, 318, 410, 420, 520] and the many attempts to simulate or model such systems [235, 420, 408, 410, 418, 259, 399, 520, 324, 406, 316, 69, 398, 315, 137, 33, 136, 232, 165, 62, 403, 224, 307, 74, 166, 227, 284, 286, 300, 323, 3, 124, 239, 270, 293, 308, 318, 319, 356, 495, 151, 173, 172, 262, 295, 294, 296, 400, 491, 519, 521, 141, 189, 264, 41]. Many important attempts and measurements undoubtedly exist that regrettably I do not know about.

In the idealized solutions of textbooks, solutes are totally isolated and solutions exist without containers or boundaries. Nothing interacts with anything in these idealized solutions but **'*everything interacts with everything*'** in the reality of ionic mixtures of living solutions. Nothing interacts by repulsion in the family of Poisson Boltzmann theories, although of course everything interacts electrostatically as points would. Geometric shapes and interactions between ionic particles do not exist in Poisson Boltzmann, so spherical (and of course other) shapes cannot 'distort' the electric field or have excluded volume. Molecules cannot have internal motions in Poisson Boltzmann, if they can be said to exist at all.

**Ions are usually crowded where they are important.** These difficulties all acquire startling importance near electrodes in electrochemical cells or near DNA, ion channels or enzymes, where ions are crowded together sometimes to the exclusion of water. Indeed, Poisson Boltzmann theories fail most dramatically to describe ions in just those places where ions are most important, near and in the structures that use ions to control or perform macroscopic functions.

Recently, these problems have been noticed by mathematicians and a number of approaches have been tried in papers I know of [335, 91, 92, 523, 524, 37, 268, 266, 155, 423, 422, 198, 426, 200, 191, 425]. (No doubt many papers have, to my regret, escaped my attention.) These approaches differ in many ways and it is far too early to choose among them. In my opinion, the correct model and the correct mathematics to implement that model are both unknown. All the models and methods must be tested against actual experimental data before scientists and mathematicians can choose intelligently among them. Fortunately, a great deal of experimental data has been available for a very long time, and new techniques are providing new data all the time. The data best known to me concern the flow of ions through proteins called ion channels [102, 103, 105, 104, 226] that control a large fraction of life.

**Ion channels are among most important devices in biology**. Channel proteins are natural nanovalves that control the flow of ions and thus a wide range of biological phenomena [244, 159, 158, 141, 147]. The following description of ion channels is adapted from an article I have written for the Springer *Encyclopedia of Applied Electrochemistry* (editors: Savinell, Ota, Kreysa) that will appear in the next year or two.

**Ion channels are proteins with holes down their middle** that control the flow of ions and electric current across otherwise impermeable biological membranes [21, 359, 246, 435]. The flow of $Na^+$, $K^+$, $Ca^{2+}$, and $Cl^-$ ions have been central issues in biology for more than a century. The flow of current is responsible for the signals of the nervous system that propagate over long distances (meters). The concentration of $Ca^{2+}$ is a 'universal' signal that controls many different systems inside cells. The concentration of $Ca^{2+}$ and other messenger ions has a role in life rather like the role of the voltage in different wires of a computer. Ion channels also help much larger solutes (e.g., organic acid and bases; perhaps polypeptides) to cross membranes but much less is known about these systems.

Ion channels can select and control the movement of different types of ions because the holes in channel proteins are a few times larger than the (crystal radii of the) ions themselves. Biology uses ion channels as selective valves to control flow and thus concentration of crucial chemical signals. For example, the concentration of $Ca^{2+}$ ions determines whether muscles contract or not. Ion channels have a role in biology similar to the role of transistors in computers and technology[147]. Ion channels control concentrations important to life the way transistors control voltages important to computers.

Specifically, the selectivity of the ryanodine receptor of cardiac and skeletal muscle can be understood with a model with less than a dozen parameters that never change value [191]. Detailed properties of the current through the channel were successfully predicted (in quantitative detail, with errors of a few per cent) with this model, often before the experiments were performed. Predictions were successful after drastic mutations, and in many (>100) solutions, of widely varying composition.

The calcium channel of cardiac muscle has a complex pattern of binding of ions that can be understood (over four orders of magnitude of concentration in many types of solutions, containing $Na^+$, $K^+$, $Rb^+$, $Cs^+$, $Ca^{2+}$, $Ba^{2+}$, $Mg^{2+}$, and so on) with a model containing two or three parameters [59]. Specific mutations change this model of a calcium channel into a sodium channel [58] just as the mutations change the selectivity in experiments. A reduced model can explain these data. The model has just three parameters that never change value, namely the diameter of the channel, the dielectric coefficient of the solution, and the dielectric coefficient of the protein. Ion diameters in the model never change value. Reduced models of this type have not yet accounted for the selectivity of potassium channels and it is not clear if they can. Reduced models are not alone in this failure. High resolution models of potassium channels are also unable to compute selectivity under a range of concentrations resembling those found in experiments and biological systems. References to this immense literature can be found in [1]

**Simulations of reduced models [491] and analysis [377, 421] give a feel for how far one can go** with simple models of hard spheres in a dielectric in physical systems. Gillespie's work on ion channels [191] shows how far one can go in biological channels that use flow (see his Supplementary material), along with [58, 57, 59] reviewed in [141].

**Biology can be easier than chemistry.** In many cases, biology can be described by reduced models that do not work in ionic solutions in general. The reason seems clear. Biological systems often evolve to have simple robust properties so the systems can interact with other biological systems in a reproducible way. Life consists of modules connected together into larger modules, and then into larger modules, again, eventually making organisms, and even populations of organisms. It is hard to see how life could exist if the modules themselves did not have robust reasonably simple properties under natural conditions. Indeed, engineers take the same approach as evolution, although one uses logic (and the marketplace) and the other uses natural selection of random mutants (its own kind of marketplace). Engineering systems are designed to follow robust and simple rules. A collection of transistors connected randomly have no simple description. The same transistors connected with a definite structure can form an amplifier, with a robust and simple description. The amplifier multiplies its input by a constant. A specific and complex structure is needed to perform this simple function in engineering. A specific and complex structure is needed to perform simple functions in biology.

**Both engineers and evolution use complex structures to make interacting physical laws execute simple robust behavior necessary for machines or life.**

<u>The most striking success of these reduced models</u> has been the description of the main calcium channel that controls contraction in muscle (and appears in neurons and many other cell types as well, so far with unknown function), the Ryanodine Receptor RyR [164, 278, 371, 80, 76, 77, 78, 169, 370, 201, 497, 389].

　　Gillespie has shown that an extension of the reduced model called PNP-DFT [198, 199, 200] does remarkably well [191, 193, 196, 197, 314, 313] in predicting experiments of some complexity and subtlety (i.e., anomalous mole fraction effects and three cation mixtures) as well as drastic mutations changing charge densities from some 13 molar to zero. Specifically, the selectivity of the ryanodine receptor of cardiac and skeletal muscle can be understood with a model with less than a dozen parameters that never change value [191]. Detailed properties of the current through the channel were successfully predicted (in quantitative detail, with errors of a few per cent) with this model, often before the experiments were performed. Predictions were successful after drastic mutations, and in many—more than one hundred— solutions, of widely varying composition.

　Many of these predictions were made before the experiments were done. Other experiments had been done previously, for work with a related but unsatisfactory reduced model that did not deal correctly with the impenetrability of bio-ions [80, 77, 78, 76]. That theory was developed [198, 199, 200] before variational methods were known to me, and is not self-consistent.

<u>Reduced models have been applied</u> to other channels. The calcium channel of cardiac muscle has a complex pattern of binding of ions that can be understood (over four orders of magnitude of concentration in many types of solutions, containing $Na^+$, $K^+$, $Rb^+$, $Cs^+$, $Ca^{2+}$, $Ba^{2+}$, $Mg^{2+}$, and so on) with a model containing two or three parameters [59]. Specific mutations change this model of a calcium channel into a sodium channel [58] just as the mutations change the selectivity in experiments. A reduced model can explain these data. The model has just three parameters that never change value, namely the diameter of the channel, the dielectric coefficient of the solution, and the dielectric coefficient of the protein. Ion diameters in the model never change value. Reduced models of this type have not yet accounted for the selectivity of potassium channels and it is not clear if they can.

The original publications [155, 266] computed curves of binding selectivity in two classical channels of considerable biological interest the calcium channel of cardiac muscle [230, 310, 229, 9, 243, 366, 238, 367, 163, 365, 440, 59, 154, 55, 313] and the voltage activated sodium channel of nerve [247, 248, 22, 49, 45, 47, 110, 46, 401]. Both were represented by a reduced model of the protein in which side chains are represented as spheres free to move within the channel selectivity filter, but not able to move out of that region [58, 57, 59, 55, 221, 153, 141].

This model has proven remarkably successful in dealing with the important selectivity phenomena of these channels. A single model, with one set of parameters that are never changed, using crystal radii of ions, and one dielectric coefficient and one channel diameter, is able to account for selectivity data in a wide range of solutions (over 4 orders of magnitude of calcium concentration, and in solutions of varying $K^+$, $Na^+$, $Rb^+$ and $Cs^+$ concentration, for example). The calcium channel is represented by side chains Glu Glu Glu Glu and the sodium channel by Asp Glu Lys Ala. This work is reported in some 35 papers using various numerical methods including the variational model described here.

**Inverse Problem has been solved.** It has been possible to invert the analysis of reduced models of the calcium channel. The inverse problem of determining the distribution of side chains inside the channel from current voltage relations in a range of solutions has actually been explicitly solved, using established methods of inverse problems, including the effects of noise and systematic error [66]. In fact, the inverse problem was well-posed, with the problem being too much data, not enough, in contrast to the forward problem, where current voltage curves predicted by these models are exquisitely (and frighteningly) sensitive to tiny changes in the model, including tiny changes in diameter or structure[154, 141, 221]. In the forward problem, the structure is so important that it must be a computed consequence of the forces, an output of analysis, if results are to be reliable. Simulations must compute these structures (and the distribution of these structures) reliably if predictions of current voltage relations are to be robust.

**Inconsistencies in DFT-PNP**. The DFT-PNP method [198, 199, 200, 66, 193, 194, 197] model is built on an approximate treatment of ions in water [198, 199, 200] in which the excess free energy of a system without flow is simply added to the classical free energy of the PNP model. It does not deal explicitly, if at all, with effects of the electric field (e.g., relaxation and dielectrophoresis effects) identified and studied in classical work on ionic conductance [179, 177, 180, 186, 178, 181, 183, 184, 185, 182, 187, 287], even in textbooks [321]. DFT-PNP also assumes local equilibrium, as do other approaches using combinations of simulation and PNP equations [121, 54, 432]. Assumptions like these are reasonable in and near a channel protein, where the ionic atmosphere is dominated by the protein itself and flows are not too large. (It is interesting that those deviations that are found between theory and data for Gillespie's model of the ryanodine receptor ([191], see Supplementary Material) seem to occur mostly when flows are large and the assumption of local equilibrium is in danger.)

For bulk solutions in general, the lack of self-consistency in PNP-DFT prevents it from being automatically accepted as a good general model of ionic solutions in my view [287]. Use of DFT-PNP for bulk solutions, will remain problematic in my view, until it is shown to agree with classical results and analysis of ionic conductance in bulk solutions [179, 177, 180, 186, 178, 181, 183, 184, 185, 182, 187, 287].

**Local equilibrium assumed in DFT-PNP is inconsistent with global flow.** I am concerned about the assumption of local equilibrium in DFT-PNP (applied to bulk solutions) because it is not consistent with the existence of flux. It must be clearly understood that any assumption of local equilibrium is also an assumption of local zero flux. It is not clear how a system can have zero local flux and long range substantial flux, particularly when the system is a nanovalve connected in series with a high impedance entry process, and macroscopic baths. It is not clear that the tension between incompatible assumptions can be resolved in a unique way giving well posed predictions to compare with experiments. Theories with ambiguous predictions can create significant confusion in science that takes decades to resolve.

An important advantage of the energy variational methods discussed later in this paper [342, 456, 522, 132, 345, 155, 269, 266, 524, 523, 382, 433] is their precise definition as part of the theory of complex fluids. These methods provide unique results because the mathematics is unique. The methods are, for example, indifferent to flow. The variational treatment of ionic conductance [267, 382, 266, 269, 155] makes no simplifying assumptions about local equilibrium: everything (that is in the model) interacts with everything else, to minimize the dissipation and (Helmholtz free energy) of the model. The methods work when flows are vigorous and when they are zero, at thermodynamic equilibrium. Thus, calculations can be done in the nonequilibrium situations and mixed ionic solutions used nearly always in experimental work.

Assumptions of local equilibrium have been made before, e.g., in the theory of Brownian motion over potential barriers, Kramers' problem [162, 170, 231, 306, 304, 311, 386, 387, 388, 393, 409, 445, 444, 443, 446, 448, 447, 449, 465, 485]. Subtleties and near paradoxes [170, 231, 485] were not resolved for years, until a self-consistent analysis was performed [162, 305]. That analysis took several years because we had to deal with a <u>second order</u> Langevin equation (with sets of <u>doubly conditioned</u> trajectories [162, 305] corresponding to the unidirectional fluxes of macroscopic biophysics) even though friction dominated the problems. Previous treatments (for example, those of Smoluchowski, Langevin, and Einstein) of a first order stochastic differential equation had to be replaced with a second order stochastic equation so a pair of boundary conditions could be satisfied. Those equations were solved using a singular perturbation analysis to introduce the high friction case into the resulting Fokker-Planck equations while preserving nonzero flow. A pair of boundary conditions was an absolute requirement in our view if we were to account for macroscopic diffusion, as described by Fick, in which ions move from one concentration to another and thus satisfy two boundary conditions. The classical treatments of Smoluchowski, Langevin, and Einstein use first order Langevin equations which can only satisfy one boundary condition and thus cannot account for the classical phenomena of macroscopic diffusion. The asymptotic analysis used previously implied (nearly) zero flow, and thus did not address the problem we were interested in, namely Fick's problem of macroscopic flow between regions of different concentration.

With this stochastic analysis, the paradoxes of flow at equilibrium resolved into the simplicity of the classical theory of mass action [150]. Much to the surprise of the authors, classical theory of the theory of mass action turned out to survive this detailed stochastic analysis of the underlying Kramers' problem. Diffusion between two concentrations as in Fick's original treatment of macroscopic diffusion could be written exactly as rate equations of the theory of mass action. The classical theory was preserved in form, but the rate constants of the diffusion reaction acquired more precise meaning and explicit expressions for rate constants for diffusion were derived. The rate constants were variables, not constants. Thus, classical work using rate formulations for diffusion were justified, under one set of conditions. But if conditions were changed, as is almost always the case in experiments and applications, rate constants had to be changed, if the theory was to describe diffusion over a barrier, i.e., Kramers problem. Solving this Kramers problem in a self-consistent way took a number of years. Replacing the inconsistent assumption of local equilibrium allowed derivation of a simple and powerful representation of diffusion as a chemical reaction.

Returning now to DFT-PNP, I make the obvious conclusion. The difficulties in DFT-PNP produced by the assumption of local equilibrium (inconsistent with the existence of global flow) are hard to foresee and are likely to be even harder to resolve. The complex phenomena of classical conduction theory (best described in Justice [287], in my view) may not all be present in DFT-PNP with its inconsistencies. DFT-PNP depends on physical approximations that were the best that could be done at the time [199, 198], but those approximations cannot substitute for mathematics, in my view, in the general case of bulk solutions.

I think it preferable to use methods that are inherently self-consistent and do not contain the possibility of internal paradox. An important advantage of the energy variational methods discussed below [342, 456, 522, 132, 345, 155, 269, 266, 524, 523, 382, 433] is their automatic extension of equilibrium operators to nonequilibrium situations (produced for example by spatially non-uniform boundary conditions) in a mathematically precise and defined way, always fully self-consistent.

**Molecular Dynamics Simulations of Proteins.** Proteins allow atomic scale structures to control macroscopic function so it is natural to seek understanding with models that include all atomic detail using the methods of molecular dynamics to compute atomic motion.

Simulations of molecular dynamics by themselves so far have not helped provide understanding of equilibrium let alone flow of living solutions, mixtures of mono- and divalents. Simulations of molecular dynamics do not describe the activity—let alone flows—of mixtures of ions in biological solutions that are always mixtures where calcium is usually important. Simulations have grave difficulties dealing with the range of concentrations that are controlled in biological experiments (roughly, $10^{-10}$ to 1 M). ***By themselves*** it seems that simulations can never solve these problems because of the multiscale issues that must be dealt with ***all at once.*** Interactions of so many types across so many scales are just too much [151, 141] to deal with numerically. Indeed, interactions are so strong, variable, and subtle that even the definition of the properties of single ions (hypothetically non-interacting) is a daunting task taking 664 pages and 2406 references [264] in a recent monograph from leading workers in the field.

It is hard to know how one could even write force fields (of the type used in classical or polarizable molecular dynamics) that would work in the concentrated environments near and in DNA, enzyme active sites, or ionic channels [480, 340, 261]. I remind colleagues that such force fields and simulations must actually calculate the activities of ionic solutions and mixtures correctly. It is not good enough to calculate them incorrectly or in some idealized infinitely dilute solutions. Biological molecules require bio-ions (that accompany the molecules in Ringer solutions, for example) to be of the right type and in the right amount (i.e., concentrations). Otherwise, the biological molecules do not work, or even die ('denature' or 'inactivate' in biological language.) These are the solutions used by experimental colleagues and they must be calculated correctly if the simulations of molecular dynamics are to deal with experiments as they are actually done.

A major problem facing molecular dynamics is that atoms move a great deal, at the speed of sound, as a first approximation [43], and so computations must resolve $10^{-16}$ sec. Little biology happens faster than $10^{-4}$ seconds. Atomic scale computations must extend over 12 orders of magnitude in time if they are to compute biological function.

Current flow through channels depends on bulk concentrations of ions ranging from $10^{-7}$ M (or even much less) to 0.5 M. Concentrations of ions in and near ion channels (and in and near active sites of enzymes) are much higher, even larger than 10 M because the systems are so small and have large densities of permanent charge from acid and base side chains of proteins. Computations of these concentrations must include nonideal properties of highly concentrated interacting solutions and side chains because these are known to be of great importance in ionic mixtures or solutions greater than 50 mM concentration. Simulations in full atomic detail must also include staggering numbers of water molecules to deal with the trace concentrations of $10^{-11}$ to $10^{-7}$ M of signaling ions, e.g., $Ca^{2+}$.

Molecular dynamics of biological function must be done in nonequilibrium conditions where flows occur, because almost all biology occurs in these conditions. Simulations have difficulty dealing with the action potentials of nerve and muscle fibers. Molecular dynamics cannot compute the billions of trajectories of ions that cross membranes to make action potentials, lasting milliseconds to nearly a second, flowing centimeters to meters down nerve axons. Simulations at present cannot deal with flows that are controlled by channels on the atomic scale but couple to boundary conditions on the macroscopic scale, millimeters away from the channel. Many biological systems use atomic scale structures this way to control macroscopic function.

**Simulations must deal with all issues of scale at once,** because biology uses them all at once. For these reasons, molecular dynamics cannot deal directly with biological function, as of now. A multiscale approach with explicit models and calibrated links between scales seems unavoidable, as in the analysis of the propagating voltage signal of nerve fibers, and in engineering technology [141], in general.

**Multiscale reduced models of ion channels are feasible.** Reduced models of some open ionic channels have proven surprisingly successful [141] as previously discussed. In several cases, it has been possible to understand, and predict experimental results before they were done. Channels have been built that behave as expected [496].

**Successful reduced models include surprisingly little atomic detail;** for example, they treat water as a continuum dielectric and side chains as spheres. It is not clear why a sensitive biological function like selectivity can be explained with such little regard to the atomic details of hydration and solvation, but the evidence is clear. In several important cases, the explanation is successful.

**Experiments depend sensitively on some variables but not on others.** Experiments on ion channels are designed for a reason and typically depend sensitively on some variables but not on others. Even the name of ionic channels depends on the identification of their 'reversal' potential' with the gradient of chemical potential of a particular ion. Simulations must calculate the chemical potential of ions correctly in the bulk if ion channels are to be named correctly. Accuracy of at least ±5 mV is needed because the thermal energy of biological diffusion is $k_B T / e = 25\text{mV}$. Simulations must calculate the chemical potential of ions correctly in and near enzyme active sites, ion channels, and binding proteins if the function of these biological systems are to be understood. One step would be to show that molecular dynamics simulations correctly calculate the chemical potential of ionic mixtures near boundaries of physical systems [262].

**Biological reality must determine the choice of mathematical treatment.** Certainly, the mathematical treatment of ionic solutions in biology must deal with mixtures of different ions of widely different concentrations. Biologists typically deal with concentrations from molar to nanomolar, or even smaller, in their daily experiments.

Certainly, the mathematical treatment of ionic solutions must deal with flows [61]. Gradients and flows are used in biology to create the devices and machines of life, rather as they are in engineering. Engineering devices are hardly worth studying when their power supplies are turned off and their simple device laws no longer hold true. They also are far more complex and sensitive to irrelevant conditions, when they are at thermodynamic equilibrium, no longer a robust device. Living systems are much less interesting when they are dead, whether those systems are corpses or crystallized proteins. One cannot expect living systems to be the same devices when dead as they are when alive. Mathematics must describe biological reality and experiments as they actually occur.

Biology occurs in complex ionic system in which everything interacts with everything else. It seems obvious that such interacting systems need to be analyzed by mathematics designed to handle interactions, not be mathematics designed to handle ideal noninteracting systems (like the isolated atoms of perfect gases of thermodynamics). The mathematics of interacting systems must always be self-consistent. Otherwise, descriptions of one set of conditions can only work under that set of conditions. When any condition changes, everything changes (if everything interacts with everything else), and so all interactions change. There seems no hope of catching such changes in interactions in a sequence or composite of non-interacting models. Rather, one needs a mathematical structure that deals with interactions automatically and self-consistently. In such a structure, interactions automatically change (as required and specified by mathematics) when any one physical condition or component is changed.

**Mathematics must describe biological experiments** as they are actually done. Scientists cope with complex systems by *simplifying systems and then adding back components or fields, one by one.* It is difficult to describe this hierarchy of systems if one deals with each system individually, without interactions, in the ideal tradition of chemistry. Laboratories use inconsistent models of the system and make different choices of parameters. It is hard to compare inconsistent models.

**A hierarchy of inconsistent models is a challenge to the scientific process.** The scientific (social) process may not converge. It is understandable, nearly inevitable in such circumstances that theories of mixtures of electrolytes (e.g., equations of state) should include large numbers of vaguely defined parameters, not useful beyond the conditions in which they were measured.

Scientists have been crippled by their lack of self-consistent mathematics, not by their lack of skill or energy, in my opinion. They did their best with the mathematical tools they had. Self-consistency can make an enormous difference. Self-consistency is an enormous help in focusing attention, and decreasing distracting discord, as it has been in computational electronics. Inconsistent models are particularly damaging in systems of ions and microelements that interact on many scales, as they have, between boundary conditions. A recent treatise ([264]: 664 pages and 2406 references) shows what happens when a system in which everything interacts with everything else is diligently analyzed to exhaustion without self-consistent treatment of interactions.

**Importance of self-consistent analysis** can be seen in the history of semiconductor physics and computational electronics [457, 486, 357, 228, 469, 30, 472, 451, 363, 427, 458, 279, 405, 263, 419, 120, 242, 352, 225, 168, 461, 123, 488, 487]. The analysis of semiconductor devices has always sought to be self-consistent. Interactions of point quasi-particles (holes and 'semi-electrons' [109, 149, 147]) have been treated self-consistently, along with both short and long range properties of the electric field. It is the correct treatment of interactions that has made possible the enormous progress of Moore's law [419, 457, 488, 380, 381, 351]. Without a correct mathematics of interactions, without self-consistent analysis, one must do trial and error experimentation. The methods of trial and error analysis have been wonderfully refined in the biological and pharmaceutical sciences. Nonetheless, refined as they may be, they are far from as efficient as self-consistent analysis. One cannot imagine an airplane [12], particularly a supersonic airplane, designed by trial and error analysis. Trial and error analysis cannot deal productively with highly interacting spatially varying systems with flows on all scales [11, 12, 20, 133, 283, 528, 50, 51, 190, 258, 322, 269, 434, 132, 345]. For that reason, I believe trial and error analysis of ionic and biological systems needs to be supplemented by self-consistent mathematics designed to deal with the interactions that characterize complex systems.

**Complex systems are well described by variational methods** [269, 68, 132] well known to mathematicians. Sadly, the power of these methods is not well known to biologists and chemists so I describe some of that power here, despite the danger that I am reiterating to mathematicians what they know better than I. Perhaps it is important for mathematicians to be reminded of the properties of self-consistent and variational methods that biologists and chemists do not know as well as they might.

Scientists need to replace their idealized noninteracting models with a self-consistent treatment in which everything can interact with everything else. Mathematicians working on ionic solutions [335, 526] are well aware that variational methods allow components and fields to be added or subtracted in functionals, from which differential equations are derived by the Euler-Lagrange process. Mathematicians need to spread their knowledge of variational methods to the physical chemists, physiologists, and molecular biologists of the world, Mathematicians need to help them to solve the big problems.

Self-consistent variational methods allow one to **derive** the differential equations that describe the system. It is difficult to write down such equations when many fields and components are involved without a derivation from a variational principle. It is all too easy to leave something out, including effects from different physics, described as a different field, often on different scales, or to invent many parameters that are hard to determine. It is difficult to know how to add components or fields without disturbing the other parts of a system of partial differential equations. A great deal of experimentation consists of simplifying systems and then adding back components or fields one by one. It is difficult to describe such experimental situations uniquely if one combines partial differential equations.

**Variational methods allow components and fields to be added or subtracted** in functionals, from which differential equations are derived by the Euler Lagrange process. Energy variational methods [342, 456, 522, 132, 345, 155, 269, 266, 524, 523] allow one to describe systems with energy and dissipation functionals from which partial differential equations are derived. The resulting partial differential equations are always self-consistent, if the algebra is done correctly.

**Energy Variational methods embody physics.** The variational methods I prefer—with admitted bias on my part—are not arbitrary mathematical structures. Energy variational methods aspire [132, 269] to be a natural extension of thermodynamics, joining free energy and dissipation functionals, as envisioned by Onsager and followers. These methods combine [132, 155, 382, 433] the Least Action Principle of mechanics with the Maximum Dissipation Principle of Rayleigh, later applied by Onsager, including eventual time dependent relaxation to the steady-state. The derivation of the Navier Stokes equation for incompressible flow [269] illustrates the approach.

The mathematics of variational methods is consistent with itself, but it may not be consistent with the real world. Energy variational methods are not magic. If the underlying models are incorrect or incomplete, the results of a variational analysis will be incorrect or incomplete. But the variational results are never inconsistent mathematics and they can be derived to have minimal free parameters.

**Conclusion.** I write to tell the mathematical community of biological reality and mathematical challenge. The mathematical analysis of ionic solutions is a topic of profound importance and opportunities. It looms almost too close to see.

Mathematicians can use energy variational principles to deal with the reality of charged spheres in a frictional dielectric. They can then move to describe the water and ionic correlations more and more realistically, as experiments dictate. Energy variational methods allow the systematic analysis and improvement of models of the mixed ionic solutions of life. Energy variational methods allow molecules in solutions to be microelements in a self-consistent theory of complex fluids. Classical problems of chemistry and biology can be attacked with the computational power of modern mathematics when a self-consistent variational theory is used to describe the energy and dissipation of these systems.

**The Big Picture.** The big problem requires mathematics that describes interacting electrolytes in devices. Mathematics can start with bio-ions described as hard spheres diffusing in a uniform dielectric. Such analysis is already feasible. Ionic solutions can be studied with the existing theory of complex fluids.

Work on the Poisson-Boltzmann equation [518] sets the stage on which the moving dance of biology can now be studied, as it is actually lived. Analysis of the PB equations provides the initial iterates for numerical procedures needed to solve the big problem (with all interactions). **Eventually the big problem must itself be addressed.** I think eventually can be now. It can be done if the community of mathematicians wishes to make it so.

**It will take an army of mathematicians** to study the ionic solutions of physical chemistry and biology as complex fluids. Mathematicians must learn the experimental traditions of physical chemistry and physiology before they can address long standing unsolved problems. They must rework their tools to deal with the realities of electrolytes in solutions and near channels, proteins, and electrodes. No one can know how these tools will succeed (and fail) without trying. But challenges can only be met by trying. I believe daunting interactions of ions, microelements, and the macroscopic world can be handled automatically and self-consistently by the theory of complex fluids.

Other molecules—beyond the bio-ions—will need models with more microdynamics. Reduced models of these molecules are often already known, and can be improved by appropriate extensions of the theory of inverse problems, I suspect. Chemical reactions can perhaps be treated by the theory of complex fluids, as already mentioned, as interactions of microelements (reactants) involving rearrangements of internal (electronic) structures of reactants, according to Schrödinger's wave equation of the electron.

**Challenge**. I challenge mathematicians to apply their tools and skills to the reality of ions in chemistry and biology. Mathematicians can use self-consistent theories of complex fluids to allow systematic analysis and improvement of models of the plasmas of ionic solutions, in life and in our chemical laboratories. Computations are needed of current voltage relations in complex mixtures of ions of many types and concentrations. Computations are underway for bio-ions that are nearly hard spheres. Theories of complex fluids need to be applied to classical unsolved problems of chemistry and biology, involving plasmas of bio-ions, organic compounds, proteins and nucleic acids. It can be done but not with theories of simple fluids.

## **Acknowledgement**

I thank Wei Cai for suggesting that I write this paper and Gail Corbett for much encouragement along the way.

A short version of this note has been prepared for *SIAM News*. The description of ion channels was prepared for the Springer-Verlag *Encyclopedia of Applied Electrochemistry* (editors: Savinell, Ota, Kreysa) and will appear in the next year or two. A version of this paper will be submitted by invitation to *Structural Bioinformatics,* part of the Springer series *Advances in Experimental Medicine and Biology,* editor Dongqing Wei, JiaoTong University (Shanghai). See http://www.springer.com/series/5584 http://www.springer.com/series/5584 ,.

Continual joyous interactions with Chun Liu have shaped my thoughts and this paper. I am grateful for support from the Bard Endowed Chair of Rush University that made this work possible.

**References**

[1]     *Obituary: Professor Niels J. Bjerrum*, Transactions of the Faraday Society, 55 (1959).

[2]     N. ABAID, R. S. EISENBERG and W. LIU, *Asymptotic expansions of I-V relations via a Poisson-Nernst-Planck system*, SIAM Journal of Applied Dynamical Systems, 7 (2008), pp. 1507.

[3]     Z. ABBAS, E. AHLBERG and S. NORDHOLM, *Monte Carlo Simulations of Salt Solutions: Exploring the Validity of Primitive Models*, The Journal of Physical Chemistry B, 113 (2009), pp. 5905-5916.

[4]     Z. ABBAS, M. GUNNARSSON, E. AHLBERG and S. NORDHOLM, *Corrected DebyeHuckel Theory of Salt Solutions: Size Asymmetry and Effective Diameters*, The Journal of Physical Chemistry B, 106 (2002), pp. 1403-1420.

[5]     S. ABOUD, D. MARREIRO, M. SARANITI and R. EISENBERG, *A Poisson P3M Force Field Scheme for Particle-Based Simulations of Ionic Liquids*, J. Computational Electronics, 3 (2004), pp. 117-133.

[6]     M. AGUILELLA-ARZO, V. AGUILELLA and R. S. EISENBERG, *Computing numerically the access resistance of a pore* European Biophysics Journal,, 34 (2005), pp. 314-322.

[7]     B. ALBERTS, D. BRAY, J. LEWIS, M. RAFF, K. ROBERTS and J. D. WATSON, *Molecular Biology of the Cell*, Garland, New York, 1994.

[8]     T. ALLEN, S. KUYUCAK and S. CHUNG, *Molecular dynamics estimates of ion diffusion in model hydrophobic and KcsA potassium channels.*, Biophys Chem., 86 (2000), pp. 1-14.

[9]     W. ALMERS and E. W. MCCLESKEY, *Non-selective conductance in calcium channels of frog muscle: calcium selectivity in a single-file pore.*, J.Physiol., 353 (1984), pp. 585-608.

[10]    M. D. ALTMAN, A. ALI, G. S. REDDY, M. N. NALAM, S. G. ANJUM, H. CAO, S. CHELLAPPAN, V. KAIRYS, M. X. FERNANDES, M. K. GILSON, C. A. SCHIFFER, T. M. RANA and B. TIDOR, *HIV-1 protease inhibitors from inverse design in the substrate envelope exhibit subnanomolar binding to drug-resistant variants*, J Am Chem Soc, 130 (2008), pp. 6099-113.

[11]    J. ANDERSON, *Computational Fluid Dynamics*, McGraw-Hill Science/Engineering/Math, New York, 1995.

[12]    J. ANDERSON, *Fundamentals of Aerodynamics*, McGraw-Hill Science/Engineering/Math, New York, 2007.

[13]    J. L. ANDERSON and D. W. ARMSTRONG, *Immobilized ionic liquids as high-selectivity/high-temperature/high-stability gas chromatography stationary phases*, Anal Chem, 77 (2005), pp. 6453-62.

[14]    J. ANTOSIEWICZ, M. K. GILSON, I. H. LEE and J. A. MCCAMMON, *Acetylcholinesterase: diffusional encounter rate constants for dumbbell models of ligand*, Biophys J, 68 (1995), pp. 62-8.

[15]    J. ANTOSIEWICZ, J. A. MCCAMMON and M. K. GILSON, *The determinants of pKas in proteins*, Biochemistry, 35 (1996), pp. 7819-33.

[16]    J. ANTOSIEWICZ, J. A. MCCAMMON and M. K. GILSON, *Prediction of pH-dependent properties of proteins*, J Mol Biol, 238 (1994), pp. 415-36.

[17]    J. ANTOSIEWICZ, J. A. MCCAMMON, S. T. WLODEK and M. K. GILSON, *Simulation of charge-mutant acetylcholinesterases*, Biochemistry, 34 (1995), pp. 4211-9.

[18]    K. ARNING, *Mathematical Modelling and Simulation of Ion Channels*, *Radon Institute for Computational and Applied Mathematics*, Johannes Kepler University Linz, Linz, 2009, pp. 139.

[19]    K. ARNING, M. BURGER, R. S. EISENBERG, H. W. ENGL and L. HE, *Inverse problems related to ion channels*, PAMM, 7 (2007), pp. 1120801-1120802.

[20]    V. I. ARNOLD and B. KHESIN, *Topological Methods in Hydrodynamics* Springer, New York, 1999.

[21]    F. M. ASHCROFT, *Ion Channels and Disease*, Academic Press, New York, 1999.

[22]    I. ATWATER, F. BEZANILLA and E. ROJAS, *Sodium influxes in internally perfused squid giant axon during voltage clamp*, J Physiol, 201 (1969), pp. 657-64.

[23]    I. ATWATER, F. BEZANILLA and E. ROJAS, *Time course of the sodium permeability change during a single membrane action potential*, J Physiol, 211 (1970), pp. 753-65.

[24]    N. A. BAKER and J. A. MCCAMMON, *Electrostatic interactions*, Methods Biochem Anal, 44 (2003), pp. 427-40.

[25]    V. BARCILON, *Ion flow through narrow membrane channels: Part I.*, SIAM J. Applied Math, 52 (1992), pp. 1391-1404.

[26]    V. BARCILON, D.-P. CHEN, R. S. EISENBERG and J. W. JEROME, *Qualitative properties of steady-state Poisson-Nernst-Planck systems: perturbation and simulation study*, SIAM J. Appl. Math., 57 (1997), pp. 631-648.

[27]    V. BARCILON, D. P. CHEN and R. S. EISENBERG, *Ion flow through narrow membranes channels: Part II.*, SIAM J. Applied Math, 52 (1992), pp. 1405-1425.

[28]    A. J. BARD and L. R. FAULKNER, *Electrochemical Methods: Fundamentals and Applications.* , John Wiley & Sons, New York, 2000.

[29]    J. BARKER and D. HENDERSON, *What is "liquid"? Understanding the states of matter*, Reviews of Modern Physics, 48 (1976), pp. 587-671.

[30]    J. R. BARKER and D. K. FERRY, *On the physics and modeling of small semiconductor devices--II : The very small device*, Solid-State Electronics, 23 (1980), pp. 531-544.

[31]    C. A. BARLOW, JR. and J. R. MACDONALD, *Theory of Discreteness of Charge Effects in the Electrolyte Compact Double Layer*, in P. Delahay, ed., *Advances in Electrochemistry and Electrochemical Engineering, Volume 6*, Interscience Publishers, New York, 1967, pp. 1-199.

[32]    J.-L. BARRATT and J.-P. HANSEN, *Basic concepts for simple and complex liquids*, Cambridge University Press, 2003.

[33]    J. BARTHEL, H. KRIENKE and W. KUNZ, *Physical Chemistry of Electrolyte Solutions: Modern Aspects*, Springer, New York, 1998.

[34]    L. BASS, A. BRACKEN and J. HILDEN, *Flux ratio theorems for nonstationary membrane transport with temporary capture of tracer*, J. Theor. Biol., 118 (1988), pp. 327-338.

[35]    L. BASS and A. MCNABB, *Flux ratio theorems for nonlinear membrane transport under nonstationary conditions. J. Theor. Biol.*, J. Theor. Biol., 133 (1988), pp. 185-191.

[36]    M. Z. BAZANT, M. S. KILIC, B. D. STOREY and A. AJDARI, *Towards an understanding of induced-charge electrokinetics at large applied voltages in concentrated solutions*, Advances in Colloid and Interface Science, 152 (2009), pp. 48-88.

[37]    M. Z. BAZANT, B. D. STOREY and A. A. KORNYSHEV, *Double layer in ionic liquids: overscreening versus crowding*, Physical Review Letters, 106 (2011), pp. 046102.

[38]    M. Z. BAZANT, K. THORNTON and A. AJDARI, *Diffuse-charge dynamics in electrochemical systems*, Physical Review E, 70 (2004), pp. 021506.

[39]    S. BEK and E. JAKOBSSON, *Brownian dynamics study of a multiply-occupied cation channel: application to understanding permeation in potassium channels*, Biophys J, 66 (1994), pp. 1028-1038.

[40]    A. BEN-NAIM, *Inversion of the Kirkwood Buff theory of solutions: Application to the water ethanol system*, AIP, 1977.

[41]    A. BEN-NAIM, *Molecular Theory of Solutions*, Oxford, New York, 2006.

[42]    A. BEN-NAIM, *Molecular Theory of Water and Aqueous Solutions Part II: The Role of Water in Protein Folding, Self-Assembly and Molecular Recognition* World Scientific Publishing Company, 2011.

[43]    S. BERRY, S. RICE and J. ROSS, *Physical Chemistry*, John Wiley, New York, 1963.

[44]    S. R. BERRY, S. A. RICE and J. ROSS, *Physical Chemistry*, Oxford, New York, 2000.

[45]    F. BEZANILLA, *Gating of sodium and potassium channels*, J Membr Biol, 88 (1985), pp. 97-111.

[46]    F. BEZANILLA, *Ion channels: from conductance to structure*, Neuron, 60 (2008), pp. 456-68.

[47]    F. BEZANILLA, *Single sodium channels from the squid giant axon*, Biophys J, 52 (1987), pp. 1087-90.

[48]    F. BEZANILLA, E. ROJAS and R. E. TAYLOR, *Sodium and potassium conductance changes during a membrane action potential*, J. Physiol. (London), 211 (1970), pp. 729-751.

[49]    F. BEZANILLA, J. VERGARA and R. E. TAYLOR, *Voltage clamping of excitable membranes*, in G. Ehrenstein and H. Lecar, eds., *Methods of Experimental Physics Volume 20*, Elsevier Inc. , New York, 1982, pp. 445-511.

[50]    R. B. BIRD, R. C. ARMSTRONG and O. HASSAGER, *Dynamics of Polymeric Fluids, Fluid Mechanics*, Wiley, New York, 1977.

[51]    R. B. BIRD, O. HASSAGER, R. C. ARMSTRONG and C. F. CURTISS, *Dynamics of Polymeric Fluids, Kinetic Theory* Wiley, New York, 1977.

[52]    J. BOCKRIS and A. REDDY, *Modern Electrochemistry*, Plenum Press, New York, 1970.

[53]    J. O. M. BOCKRIS and A. K. N. REDDY, *Modern Electrochemistry. Ionics.*, Plenum, New York, 1988.

[54]    D. BODA and D. GILLESPIE, *Steady-State Electrodiffusion from the Nernst–Planck Equation Coupled to Local Equilibrium Monte Carlo Simulations*, Journal of Chemical Theory and Computation, 8 (2012), pp. 824-829.

[55]    D. BODA, J. GIRI, D. HENDERSON, B. EISENBERG and D. GILLESPIE, *Analyzing the components of the free-energy landscape in a calcium selective ion channel by Widom's particle insertion method*, J Chem Phys, 134 (2011), pp. 055102-14.

[56]    D. BODA and D. HENDERSON, *The effects of deviations from Lorentz–Berthelot rules on the properties of a simple mixture*, Molecular Physics, 106 (2008), pp. 2367-2370.

[57]    D. BODA, W. NONNER, D. HENDERSON, B. EISENBERG and D. GILLESPIE, *Volume Exclusion in Calcium Selective Channels*, Biophys. J., 94 (2008), pp. 3486-3496.

[58]    D. BODA, W. NONNER, M. VALISKO, D. HENDERSON, B. EISENBERG and D. GILLESPIE, *Steric Selectivity in Na Channels Arising from Protein Polarization and Mobile Side Chains*, Biophys. J., 93 (2007), pp. 1960-1980.

[59]    D. BODA, M. VALISKO, D. HENDERSON, B. EISENBERG, D. GILLESPIE and W. NONNER, *Ionic selectivity in L-type calcium channels by electrostatics and hard-core repulsion*, J. Gen. Physiol., 133 (2009), pp. 497-509.

[60]    L. BOLTZMANN, *Lectures on Gas Theory*, University of California, Berkely CA, 1964.

[61]    W. BORON and E. BOULPAEP, *Medical Physiology*, Saunders, New York, 2008.

[62]    M. BOSTRÖM, F. W. TAVARES, D. BRATKO and B. W. NINHAM, *Specific Ion Effects in Solutions of Globular Proteins: Comparison between Analytical Models and Simulation*, The Journal of Physical Chemistry B, 109 (2005), pp. 24489-24494.

[63]    B. BREYER and H. H. BAUER, *Alternating Current Polarography and Tensammetry Interscience*, Interscience, New York, 1963.

[64]    C. L. BROOKS, M. KARPLUS and B. M. PETTITT, *Proteins: A Theoretical Perspective of Dynamics, Structure and Thermodynamics.*, John Wiley & Sons, New York, 1988.

[65]     R. P. BUCK, *Kinetics of bulk and interfacial ionic motion: microscopic bases and limits for the nernst—planck equation applied to membrane systems*, Journal of Membrane Science, 17 (1984), pp. 1-62.

[66]     M. BURGER, R. S. EISENBERG and H. ENGL, *Inverse Problems Related to Ion Channel Selectivity*, SIAM J Applied Math, 67 (2007), pp. 960-989

[67]     A. BURYKIN, M. KATO and A. WARSHEL, *Exploring the origin of the ion selectivity of the KcsA potassium channel*, Proteins, 52 (2003), pp. 412-26.

[68]     S. BUYUKDAGLI, M. MANGHI and J. PALMERI, *Variational approach for electrolyte solutions: From dielectric interfaces to charged nanopores*, Physical Review E, 81 (2010), pp. 041601.

[69]     H. CABEZAS and J. P. O'CONNELL, *Some uses and misuses of thermodynamic models for dilute liquid solutions*, Industrial & Engineering Chemistry Research, 32 (1993), pp. 2892-2904.

[70]     A. E. CARDENAS, R. D. COALSON and M. G. KURNIKOVA, *Three-Dimensional Poisson-Nernst-Planck Studies. Influence of membrane electrostatics on Gramicidin A Channel Conductance*, Biophysical Journal, 79 (2000).

[71]     C. E. CHANG, W. CHEN and M. K. GILSON, *Ligand configurational entropy and protein binding*, Proc Natl Acad Sci U S A, 104 (2007), pp. 1534-9.

[72]     C. E. CHANG and M. K. GILSON, *Free energy, entropy, and induced fit in host-guest recognition: calculations with the second-generation mining minima algorithm*, J Am Chem Soc, 126 (2004), pp. 13156-64.

[73]     C. E. CHANG and M. K. GILSON, *Tork: Conformational analysis method for molecules and complexes*, J Comput Chem, 24 (2003), pp. 1987-98.

[74]     J. CHE, J. DZUBIELLA, B. LI and J. A. MCCAMMON, *Electrostatic free energy and its variations in implicit solvent models*, J Phys Chem B, 112 (2008), pp. 3058-69.

[75]     D. CHEN, R. EISENBERG, J. JEROME and C. SHU, *Hydrodynamic model of temperature change in open ionic channels*, Biophysical J., 69 (1995), pp. 2304-2322.

[76]     D. CHEN, L. XU, A. TRIPATHY, G. MEISSNER and B. EISENBERG, *Selectivity and Permeation in Calcium Release Channel of Cardiac Muscle: Alkali Metal Ions*, Biophysical Journal, 76 (1999), pp. 1346-1366.

[77]     D. CHEN, L. XU, A. TRIPATHY, G. MEISSNER and R. EISENBERG, *Permeation through the calcium release channel (CRC) of cardiac muscle*, Biophys. J., 72 (1997), pp. A108.

[78]     D. CHEN, L. XU, A. TRIPATHY, G. MEISSNER and R. EISENBERG, *Permeation through the calcium release channel of cardiac muscle*, Biophys. J., 73 (1997), pp. 1337-1354.

[79]     D. CHEN, L. XU, A. TRIPATHY, G. MEISSNER and R. EISENBERG, *Rate Constants in Channology*, Biophys. J., 73 (1997), pp. 1349-1354.

[80]     D. P. CHEN, *Nonequilibrium thermodynamics of transports in ion channels*, in M. Sokabe, A. Auerbach and F. Sigworth, eds., *Progress of Cell Research: Towards Molecular Biophysics of Ion Channels*, Elsevier, Amsterdam, 1997, pp. 269-277.

[81]     D. P. CHEN, V. BARCILON and R. S. EISENBERG, *Constant field and constant gradients in open ionic channels*, Biophys J, 61 (1992), pp. 1372–1393.

[82]     D. P. CHEN and R. S. EISENBERG, *Charges, currents and potentials in ionic channels of one conformation*, Biophys. J, 64 (1993), pp. 1405–1421.

[83]     D. P. CHEN, J. LEAR and R. S. EISENBERG, *Permeation through an open channel. Poisson-Nernst-Planck theory of a synthetic ionic channel.*, Biophys. J., 72 (1997), pp. 97-116.

[84]     W. CHEN, C. E. CHANG and M. K. GILSON, *Calculation of cyclodextrin binding affinities: energy, entropy, and implications for drug design*, Biophys J, 87 (2004), pp. 3035-49.

[85]    W. CHEN, C. E. CHANG and M. K. GILSON, *Concepts in receptor optimization: targeting the RGD peptide*, J Am Chem Soc, 128 (2006), pp. 4675-84.

[86]    W. CHEN and M. K. GILSON, *ConCept: de novo design of synthetic receptors for targeted ligands*, J Chem Inf Model, 47 (2007), pp. 425-34.

[87]    W. CHEN, J. HUANG and M. K. GILSON, *Identification of symmetries in molecules and complexes*, J Chem Inf Comput Sci, 44 (2004), pp. 1301-13.

[88]    X. CHEN, Y. LIN and M. K. GILSON, *The binding database: overview and user's guide*, Biopolymers, 61 (2001), pp. 127-41.

[89]    X. CHEN, Y. LIN, M. LIU and M. K. GILSON, *The Binding Database: data management and interface design*, Bioinformatics, 18 (2002), pp. 130-9.

[90]    X. CHEN, M. LIU and M. K. GILSON, *BindingDB: a web-accessible molecular recognition database*, Comb Chem High Throughput Screen, 4 (2001), pp. 719-25.

[91]    Z. CHEN, N. A. BAKER and G. W. WEI, *Differential geometry based solvation model I: Eulerian formulation*, J Comput Phys, 229 (2010), pp. 8231-8258.

[92]    Z. CHEN, N. A. BAKER and G. W. WEI, *Differential geometry based solvation model II: Lagrangian formulation*, J Math Biol (2011).

[93]    A. CHHIH, O. BERNARD, J. M. G. BARTHEL and L. BLUM, *Transport Coefficients and Apparent Charges of Concentrated Electrolyte Solutions: Equations for Practical Use*, Ber. Bunsenges. Phys. Chem., 98 (1994), pp. 1516-1525.

[94]    S. W. CHIU, E. JAKOBSSON and H. L. SCOTT, *Combined Monte Carlo and molecular dynamics simulation of hydrated lipid-cholesterol lipid bilayers at low cholesterol concentration*, Biophys J, 80 (2001), pp. 1104-14.

[95]    S.-H. CHUNG, T. ALLEN, M. HOYLES and S. KUYUCAK, *Permeation of ions across the potassium channel: Brownian dynamics studies*, Biophysical Journal, 77 (1999), pp. 2517-2533.

[96]    S.-H. CHUNG, M. HOYLES, T. ALLEN and S. KUYUCAK, *Study of ionic currents across a model membrane channel using Brownian dynamics*, Biophysical Journal, 75 (1998), pp. 793-809.

[97]    S. CHUNG and S. KUYUCAK, *Predicting channel function from channel structure using Brownian dynamics simulations*, Clin Exp Pharmacol Physiol., 28 (2001), pp. 89-94.

[98]    S. H. CHUNG, T. W. ALLEN and S. KUYUCAK, *Conducting-State Properties of the KcsA Potassium Channel from Molecular and Brownian Dynamics Simulations*, Biophys J, 82 (2002), pp. 628-45.

[99]    S. H. CHUNG, T. W. ALLEN and S. KUYUCAK, *Modeling diverse range of potassium channels with Brownian dynamics*, Biophys J, 83 (2002), pp. 263-77.

[100]   S. H. CHUNG and B. CORRY, *Conduction properties of KcsA measured using brownian dynamics with flexible carbonyl groups in the selectivity filter*, Biophys J, 93 (2007), pp. 44-53.

[101]   R. D. COALSON and M. G. KURNIKOVA, *Poisson-Nernst-Planck theory approach to the calculation of current through biological ion channels*, IEEE Trans Nanobioscience, 4 (2005), pp. 81-93.

[102]   E. C. CONLEY, *The Ion Channel Facts Book. I. Extracellular Ligand-gated Channels*, Academic Press, New York, 1996.

[103]   E. C. CONLEY, *The Ion Channel Facts Book. II. Intracellular Ligand-gated Channels*, Academic Press, New York, 1996.

[104]   E. C. CONLEY and W. J. BRAMMAR, *The Ion Channel Facts Book III: Inward Rectifier and Intercellular Channels* Academic Press, New York, 2000.

[105]   E. C. CONLEY and W. J. BRAMMAR, *The Ion Channel Facts Book IV: Voltage Gated Channels*, Academic Press, New York, 1999.

[106]   B. E. CONWAY, *Electrochemical Data*, Greenwood Press Publishers, Westport CT USA, 1969.

[107]  K. COOPER, E. JAKOBSSON and P. WOLYNES, *The theory of ion transport through membrane channels.*, Prog. Biophys. Molec. Biol., 46 (1985), pp. 51–96.

[108]  K. E. COOPER, P. Y. GATES and R. S. EISENBERG, *Diffusion theory and discrete rate constants in ion permeation.*, J. Membr. Biol., 109 (1988), pp. 95–105.

[109]  K. E. COOPER, P. Y. GATES and R. S. EISENBERG, *Surmounting barriers in ionic channels*, Quarterly Review of Biophysics, 21 (1988), pp. 331–364.

[110]  A. M. CORREA, R. LATORRE and F. BEZANILLA, *Ion permeation in normal and batrachotoxin-modified Na+ channels in the squid giant axon*, J Gen Physiol, 97 (1991), pp. 605-25.

[111]  B. CORRY, T. W. ALLEN, S. KUYUCAK and S. H. CHUNG, *Mechanisms of permeation and selectivity in calcium channels*, Biophys J, 80 (2001), pp. 195-214.

[112]  B. CORRY, T. W. ALLEN, S. KUYUCAK and S. H. CHUNG, *A model of calcium channels*, Biochim Biophys Acta, 1509 (2000), pp. 1-6.

[113]  B. CORRY and S.-H. CHUNG, *Influence of protein flexibility on the electrostatic energy landscape in gramicidin A*, European Biophysics Journal, 34 (2005), pp. 208-216.

[114]  B. CORRY and S. H. CHUNG, *Mechanisms of valence selectivity in biological ion channels*, Cell Mol Life Sci, 63 (2006), pp. 301-15.

[115]  B. CORRY, M. HOYLES, T. W. ALLEN, M. WALKER, S. KUYUCAK and S. H. CHUNG, *Reservoir boundaries in Brownian dynamics simulations of ion channels*, Biophys J, 82 (2002), pp. 1975-84.

[116]  B. CORRY, S. KUYUCAK and S. H. CHUNG, *Dielectric self-energy in poisson-boltzmann and poisson-nernst-planck models of ion channels*, Biophys J, 84 (2003), pp. 3594-606.

[117]  B. CORRY, S. KUYUCAK and S. H. CHUNG, *Test of Poisson-Nernst-Planck theory in ion channels*, J Gen Physiol, 114 (1999), pp. 597-9.

[118]  B. CORRY, S. KUYUCAK and S. H. CHUNG, *Tests of continuum theories as models of ion channels. II. Poisson-Nernst-Planck theory versus brownian dynamics*, Biophys J, 78 (2000), pp. 2364-81.

[119]  B. CORRY, T. VORA and S. H. CHUNG, *Electrostatic basis of valence selectivity in cationic channels*, Biochim Biophys Acta, 1711 (2005), pp. 72-86.

[120]  D. L. CRITCHLOW, *MOSFET Scaling-The Driver of VLSI Technology*, Proceedings of the IEEE, 87 (1999), pp. 659-667.

[121]  E. CSANYI, D. BODA, D. GILLESPIE and T. KRISTOF, *Current and selectivity in a model sodium channel under physiological conditions: Dynamic Monte Carlo simulations*, Biochimica et Biophysica Acta, 1818 (2012), pp. 592-600.

[122]  V. DAHIREL, M. JARDAT, J. F. DUFRECHE and P. TURQ, *How the excluded volume architecture influences ion-mediated forces between proteins*, Phys Rev E Stat Nonlin Soft Matter Phys, 76 (2007), pp. 040902.

[123]  DAMOCLES, *Damocles Web Site, IBM Research*, http://www.research.ibm.com/DAMOCLES/home.html, 2007.

[124]  B.-Y. DAN, D. ANDELMAN, D. HARRIES and R. PODGORNIK, *Beyond standard Poisson–Boltzmann theory: ion-specific interactions in aqueous solutions*, Journal of Physics: Condensed Matter, 21 (2009), pp. 424106.

[125]  L. DAVID, R. LUO and M. K. GILSON, *Ligand-receptor docking with the Mining Minima optimizer*, J Comput Aided Mol Des, 15 (2001), pp. 157-71.

[126]  M. E. DAVIS and J. A. MCCAMMON, *Electrostatics in biomolecular structure and dynamics.*, Chem. Rev., 90 (1990), pp. 509–521.

[127]  R. DE LEVIE and H. MOREIRA, *Transport of ions of one kind through thin membranes*, Journal of Membrane Biology, 9 (1972), pp. 241-260.

[128]  R. DE LEVIE and N. G. SEIDAH, *Transport of ions of one kind through thin membranes. 3. Current-voltage curves for membrane-soluble ions*, J Membr Biol, 16 (1974), pp. 1-16.

[129]  R. DE LEVIE, N. G. SEIDAH and H. MOREIRA, *Transport of ions of one kind through thin membranes. II. Nonequilibrium steady-state behavior*, J Membr Biol, 10 (1972), pp. 171-92.

[130]  E. S. DEJONG, C. E. CHANG, M. K. GILSON and J. P. MARINO, *Proflavine acts as a Rev inhibitor by targeting the high-affinity Rev binding site of the Rev responsive element of HIV-1*, Biochemistry, 42 (2003), pp. 8035-46.

[131]  G. R. DIECKMANN, J. D. LEAR, Q. ZHONG, M. L. KLEIN, W. F. DEGRADO and K. A. SHARP, *Exploration of the structural features defining the conduction properties of a synthetic ion channel*, Biophysical Journal, 76 (1999), pp. 618-630.

[132]  M. DOI, *Gel Dynamics*, Journal of the Physical Society of Japan, 78 (2009), pp. 052001.

[133]  M. DOI and S. F. EDWARDS, *The Theory of Polymer Dynamics*, Oxford University Press, New York, 1988.

[134]  B. N. DOMINY and C. L. BROOKS, *Development of a Generalized Born Model Parametrization for Proteins and Nucleic Acids*, The Journal of Physical Chemistry B, 103 (1999), pp. 3765-3773.

[135]  J. F. DUFRECHE, O. BERNARD, P. TURQ, A. MUKHERJEE and B. BAGCHI, *Ionic self-diffusion in concentrated aqueous electrolyte solutions*, Physical Review Letters, 88 (2002), pp. 095902.

[136]  S. DURAND-VIDAL, J.-P. SIMONIN and P. TURQ, *Electrolytes at Interfaces*, Kluwer, Boston, 2000.

[137]  S. DURAND-VIDAL, P. TURQ, O. BERNARD, C. TREINER and L. BLUM, *New Perspectives in Transport Phenomena in electrolytes*, Physica A, 231 (1996), pp. 123-143.

[138]  J. DZUBIELLA, J. M. SWANSON and J. A. MCCAMMON, *Coupling hydrophobicity, dispersion, and electrostatics in continuum solvent models*, Phys Rev Lett, 96 (2006), pp. 087802.

[139]  J. EDSALL and J. WYMAN, *Biophysical Chemistry*, Academic Press, NY, 1958.

[140]  S. EDWARDS, B. CORRY, S. KUYUCAK and S. H. CHUNG, *Continuum electrostatics fails to describe ion permeation in the gramicidin channel*, Biophys J, 83 (2002), pp. 1348-60.

[141]  B. EISENBERG, *Crowded Charges in Ion Channels*, Advances in Chemical Physics, John Wiley & Sons, Inc., 2011, pp. 77-223 also available at http:\\arix.org as arXiv 1009.1786v1

[142]  B. EISENBERG, *Engineering channels: Atomic biology*, Proceedings of the National Academy of Sciences, 105 (2008), pp. 6211-6212.

[143]  B. EISENBERG, *Ion channels allow atomic control of macroscopic transport*, Physica Status Solidi (c), 5 (2008), pp. 708-713.

[144]  B. EISENBERG, *Ion Channels as Devices*, Journal of Computational Electronics, 2 (2003), pp. 245-249.

[145]  B. EISENBERG, *Ionic channels in biological membranes. Electrostatic analysis of a natural nanotube.*, Contemporary Physics, 39 (1998), pp. 447 - 466.

[146]  B. EISENBERG, *Ionic Channels in Biological Membranes: Natural Nanotubes*, Accounts of Chemical Research, 31 (1998), pp. 117-125.

[147]  B. EISENBERG, *Ions in Fluctuating Channels: Transistors Alive*, Fluctuations and Noise Letters, 11 (2012), pp. 76-96  Earlier version available on http://arxiv.org/ as q-bio/0506016v2.

[148]  B. EISENBERG, *Life's Solutions are Not Ideal*, Posted on arXiv.org with Paper ID arXiv:1105.0184v1 (2011).

[149]  B. EISENBERG, *Living Transistors: a Physicist's View of Ion Channels*, available on http://arxiv.org/ as q-bio/0506016v2  24 pages (2005).

[150]  B. EISENBERG, *Mass Action in Ionic Solutions*, Chemical Physics Letters, 511 (2011), pp. 1-6.

[151]  B. EISENBERG, *Multiple Scales in the Simulation of Ion Channels and Proteins*, The Journal of Physical Chemistry C, 114 (2010), pp. 20719-20733.

[152]   B. EISENBERG, *Permeation as a Diffusion Process*, in L. J. DeFelice, ed., *Biophysics Textbook On Line "Channels, Receptors, and Transporters"* [http://www.biophysics.org/btol/channel.html#5](http://www.biophysics.org/btol/channel.html#5), Published in ArXiv as arXiv:0807.0721, 2000.

[153]   B. EISENBERG, *Proteins, Channels, and Crowded Ions*, Biophysical Chemistry, 100 (2003), pp. 507 - 517.

[154]   B. EISENBERG, *Self-organized model of selectivity*, Institute of Mathematics and its Applications, IMA University of Minnesota (2009), pp. [http://www.ima.umn.edu/2008-2009/W12.8-12.08/abstracts.html](http://www.ima.umn.edu/2008-2009/W12.8-12.08/abstracts.html) *and also* [http://arxiv.org/0906.5173](http://arxiv.org/0906.5173).

[155]   B. EISENBERG, Y. HYON and C. LIU, *Energy Variational Analysis EnVarA of Ions in Water and Channels: Field Theory for Primitive Models of Complex Ionic Fluids*, Journal of Chemical Physics, 133 (2010), pp. 104104

[156]   B. EISENBERG and W. LIU, *Poisson-Nernst-Planck systems for ion channels with permanent charges.* , SIAM Journal on Mathematical Analysis 38 (2007), pp. 1932-1966.

[157]   R. EISENBERG and D. CHEN, *Poisson-Nernst-Planck (PNP) theory of an open ionic channel*, Biophysical Journal, 64 (1993), pp. A22.

[158]   R. S. EISENBERG, *Atomic Biology, Electrostatics and Ionic Channels.*, in R. Elber, ed., *New Developments and Theoretical Studies of Proteins*, World Scientific, Philadelphia, 1996, pp. 269-357.  Published in the Physics ArXiv as arXiv:0807.0715.

[159]   R. S. EISENBERG, *Channels as enzymes: Oxymoron and Tautology*, Journal of Membrane Biology, 115 (1990), pp. 1–12.  Available on arXiv as [http://arxiv.org/abs/1112.2363](http://arxiv.org/abs/1112.2363).

[160]   R. S. EISENBERG, *Computing the field in proteins and channels.*, J. Membrane Biol., 150 (1996), pp. 1–25. Also available on http:\\arxiv.org as  arXiv 1009.2857.

[161]   R. S. EISENBERG, *From Structure to Function in Open Ionic Channels*, Journal of Membrane Biology, 171 (1999), pp. 1-24.

[162]   R. S. EISENBERG, M. M. KŁOSEK and Z. SCHUSS, *Diffusion as a chemical reaction: Stochastic trajectories between fixed concentrations.*, J. Chem. Phys., 102 (1995), pp. 1767          1780.

[163]   P. T. ELLINOR, J. YANG, W. A. SATHER, J.-F. ZHANG and R. TSIEN, *Ca2+ channel selectivity at a single locus for high-affinity Ca2+ interactions*, Neuron, 15 (1995), pp. 1121-1132.

[164]   A. S. FAIRHURST and D. J. JENDEN, *Effect of ryanodine on the calcium uptake system of skeletal muscle*, Proc Natl Acad Sci U S A, 48 (1962), pp. 807-13.

[165]   W. R. FAWCETT, *Liquids, Solutions, and Interfaces: From Classical Macroscopic Descriptions to Modern Microscopic Details*, Oxford University Press, New York, 2004.

[166]   M. V. FEDOROV and A. A. KORNYSHEV, *Ionic Liquid Near a Charged Wall: Structure and Capacitance of Electrical Double Layer*, The Journal of Physical Chemistry B, 112 (2008), pp. 11868-11872.

[167]   M. X. FERNANDES, V. KAIRYS and M. K. GILSON, *Comparing ligand interactions with multiple receptors via serial docking*, J Chem Inf Comput Sci, 44 (2004), pp. 1961-70.

[168]   S. J. FIEDZIUSZKO, I. C. HUNTER, T. ITOH, Y. KOBAYASHI, T. NISHIKAWA, S. STITZER and K. WAKINO, *Dielectric Materials, Devices, and Circuits*, IEEE Transactions on Microwave Theory and Techniques, 50 (2002), pp. 706-720.

[169]   M. FILL and J. A. COPELLO, *Ryanodine Receptor Calcium Release Channels*, Physiol Rev, 82 (2002), pp. 893-922.

[170]   G. FLEMING and P. HÄNGGI, *Activated Barrier Crossing: applications in physics, chemistry and biology*, World Scientific, River Edge, New Jersey, 1993.

[171]  W. R. FORSYTH, M. K. GILSON, J. ANTOSIEWICZ, O. R. JAREN and A. D. ROBERTSON, *Theoretical and experimental analysis of ionization equilibria in ovomucoid third domain*, Biochemistry, 37 (1998), pp. 8643-52.

[172]  D. FRAENKEL, *Monoprotic Mineral Acids Analyzed by the Smaller-Ion Shell Model of Strong Electrolyte Solutions*, The Journal of Physical Chemistry B, 115 (2010), pp. 557-568.

[173]  D. FRAENKEL, *Simplified electrostatic model for the thermodynamic excess potentials of binary strong electrolyte solutions with size-dissimilar ions*, Molecular Physics, 108 (2010), pp. 1435 - 1466.

[174]  H. L. FRIEDMAN, *A Course in Statistical Mechanics.*, Prentice Hall, Englewood Cliffs, New Jersey, 1985.

[175]  H. L. FRIEDMAN, *Electrolyte Solutions at Equilibrium*, Annual Review of Physical Chemistry, 32 (1981), pp. 179-204.

[176]  H. L. FRIEDMAN, *Ionic Solution Theory*, Interscience Publishers, New York, 1962.

[177]  R. M. FUOSS, *Properties of Electrolytic Solutions*, Chemical Reviews, 17 (1935), pp. 27-42.

[178]  R. M. FUOSS, *The Velocity Field in Electrolytic Solutions*, J Phys Chem B, 63 (1959), pp. 633-636.

[179]  R. M. FUOSS and C. A. KRAUS, *Properties of Electrolytic Solutions. IV. The Conductance Minimum and the Formation of Triple Ions Due to the Action of Coulomb Forces1*, Journal of the American Chemical Society, 55 (1933), pp. 2387-2399.

[180]  R. M. FUOSS and L. ONSAGER, *Conductance of Strong Electrolytes at Finite Dilutions*, Proc Natl Acad Sci U S A, 41 (1955), pp. 274-83.

[181]  R. M. FUOSS and L. ONSAGER, *The conductance of symmetriacl electrolytes. I. Potential of total force*, J Phys Chem B, 66 (1962), pp. 1722-1726.

[182]  R. M. FUOSS and L. ONSAGER, *The Conductance of Symmetrical Electrolytes.1a IV. Hydrodynamic and Osmotic Terms in the Relaxation Field*, J Phys Chem B, 68 (1964), pp. 1-8.

[183]  R. M. FUOSS and L. ONSAGER, *THE CONDUCTANCE OF SYMMETRICAL ELECTROLYTES. I. POTENTIAL OF TOTAL FORCE*, J Phys Chem B, 66 (1962), pp. 1722-1726.

[184]  R. M. FUOSS and L. ONSAGER, *THE CONDUCTANCE OF SYMMETRICAL ELECTROLYTES. II. THE RELAXATION FIELD*, J Phys Chem B, 67 (1963), pp. 621-628.

[185]  R. M. FUOSS and L. ONSAGER, *THE CONDUCTANCE OF SYMMETRICAL ELECTROLYTES. III. ELECTROPHORESIS*, J Phys Chem B, 67 (1963), pp. 628-632.

[186]  R. M. FUOSS and L. ONSAGER, *The Kinetic Term in Electrolytic Conductance*, J Phys Chem B, 62 (1958), pp. 1339-1340.

[187]  R. M. FUOSS, L. ONSAGER and J. F. SKINNER, *The Conductance of Symmetrical Electrolytes. V. The Conductance Equation1,2*, J Phys Chem B, 69 (1965), pp. 2581-2594.

[188]  M. FUXREITER, A. WARSHEL and R. OSMAN, *Role of active site residues in the glycosylase step of T4 endonuclease V. Computer simulation studies on ionization states*, Biochemistry, 38 (1999), pp. 9577-89.

[189]  M. B. GEE, N. R. COX, Y. JIAO, N. BENTENITIS, S. WEERASINGHE and P. E. SMITH, *A Kirkwood-Buff Derived Force Field for Aqueous Alkali Halides*, Journal of Chemical Theory and Computation (2011), pp. null-null.

[190]  P.-G. D. GENNES and J. PROST, *The Physics of Liquid Crystals*, Oxford University Press, New York, 1993.

[191]  D. GILLESPIE, *Energetics of divalent selectivity in a calcium channel: the ryanodine receptor case study*, Biophys J, 94 (2008), pp. 1169-84.

[192]  D. GILLESPIE, *A Singular Perturbation Analysis of the Poisson-Nernst-Planck System: Applications to Ionic Channels*, Molecular Biophysics and Physiology, Rush, Chicago IL, 1999, pp. 183.

[193]   D. GILLESPIE and D. BODA, *The Anomalous Mole Fraction Effect in Calcium Channels: A Measure of Preferential Selectivity*, Biophys. J., 95 (2008), pp. 2658-2672.

[194]   D. GILLESPIE, D. BODA, Y. HE, P. APEL and Z. S. SIWY, *Synthetic Nanopores as a Test Case for Ion Channel Theories: The Anomalous Mole Fraction Effect without Single Filing*, Biophys. J., 95 (2008), pp. 609-619.

[195]   D. GILLESPIE, H. CHEN and M. FILL, *Is ryanodine receptor a calcium or magnesium channel? Roles of K(+) and Mg(2+) during Ca(2+) release*, Cell Calcium, 51 (2012), pp. 427-33.

[196]   D. GILLESPIE and M. FILL, *Intracellular Calcium Release Channels Mediate Their Own Countercurrent: The Ryanodine Receptor Case Study*, Biophys. J., 95 (2008), pp. 3706-3714.

[197]   D. GILLESPIE, J. GIRI and M. FILL, *Reinterpreting the Anomalous Mole Fraction Effect. The ryanodine receptor case study.*, Biophyiscal Journal, 97 (2009), pp. pp. 2212 - 2221

[198]   D. GILLESPIE, W. NONNER and R. S. EISENBERG, *Coupling Poisson-Nernst-Planck and Density Functional Theory to Calculate Ion Flux*, Journal of Physics (Condensed Matter), 14 (2002), pp. 12129-12145.

[199]   D. GILLESPIE, W. NONNER and R. S. EISENBERG, *Density functional theory of charged, hard-sphere fluids.*, Physical Review E, 68 (2003), pp. 0313503.

[200]   D. GILLESPIE, M. VALISKO and D. BODA, *Density functional theory of the electrical double layer: the RFD functional*, Journal of Physics: Condensed Matter 17 (2005), pp. 6609-6626.

[201]   D. GILLESPIE, L. XU, Y. WANG and G. MEISSNER, *(De)constructing the Ryanodine Receptor: modeling ion permeation and selectivity of the calcium release channel*, Journal of  Physical Chemistry, 109 (2005), pp. 15598-15610.

[202]   M. K. GILSON, *The bioinformatics of molecular recognition*, J Mol Recognit, 15 (2002), pp. 1.

[203]   M. K. GILSON, *Molecular recognition databases*, Biopolymers, 61 (2001), pp. 97-8.

[204]   M. K. GILSON, *Theory of electrostatic interactions in macromolecules*, Curr Opin Struct Biol, 5 (1995), pp. 216-23.

[205]   M. K. GILSON, H. S. GILSON and M. J. POTTER, *Fast assignment of accurate partial atomic charges: an electronegativity equalization method that accounts for alternate resonance forms*, J Chem Inf Comput Sci, 43 (2003), pp. 1982-97.

[206]   M. K. GILSON, J. A. GIVEN, B. L. BUSH and J. A. MCCAMMON, *The statistical-thermodynamic basis for computation of binding affinities: a critical review*, Biophys J, 72 (1997), pp. 1047-69.

[207]   M. K. GILSON, J. A. GIVEN and M. S. HEAD, *A new class of models for computing receptor-ligand binding affinities*, Chem Biol, 4 (1997), pp. 87-92.

[208]   M. K. GILSON and B. HONIG, *Calculation of the total electrostatic energy of a macromolecular system: solvation energies, binding energies, and conformational analysis*, Proteins, 4 (1988), pp. 7-18.

[209]   M. K. GILSON and B. HONIG, *Destabilization of an alpha-helix-bundle protein by helix dipoles*, Proc Natl Acad Sci U S A, 86 (1989), pp. 1524-8.

[210]   M. K. GILSON and B. HONIG, *The dielectric constant of a folded protein*, Biopolymers, 25 (1985), pp. 2097-2119.

[211]   M. K. GILSON and B. HONIG, *The inclusion of electrostatic hydration energies in molecular mechanics calculations*, J Comput Aided Mol Des, 5 (1991), pp. 5-20.

[212]   M. K. GILSON and B. H. HONIG, *Calculation of electrostatic potentials in an enzyme active site*, Nature, 330 (1987), pp. 84-6.

[213]   M. K. GILSON and B. H. HONIG, *The dielectric constant of a folded protein*, Biopolymers, 25 (1986), pp. 2097-119.

[214]   M. K. GILSON and B. H. HONIG, *Energetics of charge-charge interactions in proteins*, Proteins, 3 (1988), pp. 32-52.

[215]   M. K. GILSON and K. K. IRIKURA, *Symmetry numbers for rigid, flexible, and fluxional molecules: theory and applications*, J Phys Chem B, 114 (2010), pp. 16304-17.

[216]   M. K. GILSON and S. E. RADFORD, *Protein folding and binding: from biology to physics and back again*, Curr Opin Struct Biol, 21 (2011), pp. 1-3.

[217]   M. K. GILSON, A. RASHIN, R. FINE and B. HONIG, *On the calculation of electrostatic interactions in proteins*, J Mol Biol, 184 (1985), pp. 503-16.

[218]   M. K. GILSON, K. A. SHARP and B. H. HONIG, *Calculating the Electrostatic Potential of Molecules in Solution: Method and Error Assessment*, Journal of Computational Chemistry, 9 (1988), pp. 327-335.

[219]   M. K. GILSON, T. P. STRAATSMA, J. A. MCCAMMON, D. R. RIPOLL, C. H. FAERMAN, P. H. AXELSEN, I. SILMAN and J. L. SUSSMAN, *Open "back door" in a molecular dynamics simulation of acetylcholinesterase*, Science, 263 (1994), pp. 1276-8.

[220]   M. K. GILSON and H. X. ZHOU, *Calculation of protein-ligand binding affinities*, Annu Rev Biophys Biomol Struct, 36 (2007), pp. 21-42.

[221]   J. GIRI, J. E. FONSECA, D. BODA, D. HENDERSON and B. EISENBERG, *Self-organized models of selectivity in calcium channels*, Phys Biol, 8 (2011), pp. 026004.

[222]   J. A. GIVEN and M. K. GILSON, *A hierarchical method for generating low-energy conformers of a protein-ligand complex*, Proteins, 33 (1998), pp. 475-95.

[223]   P. GRAF, A. NITZAN, M. G. KURNIKOVA and R. D. COALSON, *A dynamic lattice Monte Carlo model of ion transport in inhomogeneous dielectric environments: method and implementation*, Journal of Physical Chemistry B, 104 (2000), pp. 12324-12338.

[224]   A. GRATTONI, M. MERLO and M. FERRARI, *Osmotic Pressure beyond Concentration Restrictions*, The Journal of Physical Chemistry B, 111 (2007), pp. 11770-11775.

[225]   P. R. GRAY, P. J. HURST, S. H. LEWIS and R. G. MEYER, *Analysis and Design of Analog Integrated Circuits*, John Wiley, New York, 2001.

[226]   J. GRIFFITHS and C. SANSOM, *The Transporter Facts Book* Academic Press, New York, 1997.

[227]   P. GROCHOWSKI and J. TRYLSKA, *Continuum molecular electrostatics, salt effects, and counterion binding—A review of the Poisson–Boltzmann theory and its modifications*, Biopolymers, 89 (2008), pp. 93-113.

[228]   H. K. GUMMEL, *A self-consistent iterative scheme for one-dimensional steady-state transistor calculations*, IEEE Trans. Electron Devices, ED-11 (1964), pp. 445-465.

[229]   S. HAGIWARA and L. BYERLY, *Calcium channel*, Annu Rev Neurosci, 4 (1981), pp. 69-125.

[230]   S. HAGIWARA and S. NAKAJIMA, *Differences in Na and Ca spikes as examined by application of tetrodotoxin, procaine, and manganese ions*, J Gen Physiol, 49 (1966), pp. 793-806.

[231]   P. HÄNGGI, P. TALKNER and M. BOROKOVEC, *Reaction-rate theory: fifty years after Kramers.*, Reviews of Modern Physics, 62 (1990), pp. 251-341.

[232]   J.-P. HANSEN and H. LÖWEN, *Effective Interactions between Electric Double Layers* Annual Review of Physical Chemistry, 51 (2000), pp. 209-242.

[233]   J.-P. HANSEN and I. R. MCDONALD, *Theory of Simple Liquids*, Academic Press, New York, 2006.

[234]   J.-P. HANSEN and I. R. MCDONALD, *Theory of Simple Liquids*, Academic Press, New York, 1986.

[235]   H. S. HARNED and B. B. OWEN, *The Physical Chemistry of Electrolytic Solutions*, Reinhold Publishing Corporation, New York, 1958.

[236]   T. HEAD-GORDON and C. L. BROOKS, *The role of electrostatics in the binding of small ligands to enzymes*, J Phys Chem B, 91 (1987), pp. 3342-3349.

[237]   J. L. HECHT, B. HONIG, Y. K. SHIN and W. L. HUBBELL, *Electrostatic potentials near the surface of DNA: Comparing theory and experiment.*, J. Phys. Chem., 99 (1995), pp. 7782-7786.

[238]   S. H. HEINEMANN, H. TERLAU, W. STUHMER, K. IMOTO and S. NUMA, *Calcium channel characteristics conferred on the sodium channel by single mutations*, Nature, 356 (1992), pp. 441-443.

[239]   D. HENDERSON and D. BODA, *Insights from theory and simulation on the electrical double layer*, Physical Chemistry Chemical Physics, 11 (2009), pp. 3822-3830.

[240]   L. J. HENDERSON, *Blood. A Study in General Physiology* Yale University Press, New Haven, CT, 1928.

[241]   L. J. HENDERSON, *The Fitness of the Environment: an Inquiry Into the Biological Significance of the Properties of Matter*, Macmillan, New York, 1913.

[242]   K. HESS, *Advanced Theory of Semiconductor Devices*, IEEE Press, New York, 2000.

[243]   P. HESS and R. W. TSIEN, *Mechanism of ion permeation through calcium channels*, Nature, 309 (1984), pp. 453-456.

[244]   B. HILLE, *Ionic Channels of Excitable Membranes*, Sinauer Associates Inc., Sunderland, 2001.

[245]   V. HNIZDO, J. TAN, B. J. KILLIAN and M. K. GILSON, *Efficient calculation of configurational entropy from molecular simulations by combining the mutual-information expansion and nearest-neighbor methods*, J Comput Chem, 29 (2008), pp. 1605-14.

[246]   A. L. HODGKIN, *Chance and Design*, Cambridge University Press, New York, 1992.

[247]   A. L. HODGKIN, *The ionic basis of electrical activity in nerve and muscle*, Biological Reviews, 26 (1951), pp. 339-409.

[248]   A. L. HODGKIN and A. F. HUXLEY, *Currents carried by sodium and potassium ions through the membrane of the giant axon of Loligo*, J. Physiol., 116 (1952), pp. 449-472.

[249]   A. L. HODGKIN and A. F. HUXLEY, *Movement of radioactive potassium and membrane current in a giant axon*, J Physiol, 121 (1953), pp. 403-14.

[250]   A. L. HODGKIN and R. D. KEYNES, *Sodium extrusion and potassium absorption in Sepia axons*, J Physiol, 120 (1953), pp. 46P-7P.

[251]   U. HOLLERBACH, D.-P. CHEN and R. S. EISENBERG, *Two- and Three-Dimensional Poisson-Nernst-Planck Simulations of Current Flow through Gramicidin-A*, Journal of Computational Science, 16 (2002), pp. 373-409.

[252]   U. HOLLERBACH, D. CHEN, W. NONNER and B. EISENBERG, *Three-dimensional Poisson-Nernst-Planck Theory of Open Channels*, Biophysical Journal, 76 (1999), pp. A205.

[253]   U. HOLLERBACH, D. P. CHEN, D. D. BUSATH and B. EISENBERG, *Predicting function from structure using the Poisson-Nernst-Planck equations: sodium current in the gramicidin A channel.*, Langmuir, 16 (2000), pp. 5509-5514.

[254]   U. HOLLERBACH and R. EISENBERG, *Concentration-Dependent Shielding of Electrostatic Potentials Inside the Gramicidin A Channel*, Langmuir, 18 (2002), pp. 3262-3631.

[255]   B. HONIG and A. NICHOLS, *Classical electrostatics in biology and chemistry.*, Science, 268 (1995), pp. 1144-1149.

[256]   B. HONIG and K. SHARP, *Macroscopic models of aqueous solutions: biological and chemical applications*, Journal of Physical Chemistry, 97 (1993), pp. 1101-1109.

[257]   B. HONIG, K. SHARP and M. GILSON, *Electrostatic interactions in proteins*, Prog Clin Biol Res, 289 (1989), pp. 65-74.

[258]   T. Y. HOU, C. LIU and J.-G. LIU, *Multi-scale Phenomena in Complex Fluids: Modeling, Analysis and Numerical Simulations*, World Scientific Publishing Company, Singapore, 2009.

[259]   A. L. HOVARTH, *Handbook of aqueous electrolyte solutions: physical properties, estimation, and correlation methods*, Ellis Horwood,, New York, 1985.

[260]   J. J. HOWARD, G. C. LYNCH and B. M. PETTITT, *Ion and solvent density distributions around canonical B-DNA from integral equations*, The journal of physical chemistry. B, 115 (2011), pp. 547-56.

[261]   J. J. HOWARD, J. S. PERKYNS, N. CHOUDHURY and B. M. PETTITT, *An Integral Equation Study of the Hydrophobic Interaction between Graphene Plates*, Journal of Chemical Theory and Computation, 4 (2008), pp. 1928-1939.

[262]   J. J. HOWARD, J. S. PERKYNS and B. M. PETTITT, *The behavior of ions near a charged wall-dependence on ion size, concentration, and surface charge*, The journal of physical chemistry. B, 114 (2010), pp. 6074-83.

[263]   R. T. HOWE and C. G. SODINI, *Microelectronics: an integrated approach*, Prentice Hall, Upper Saddle River, NJ USA, 1997.

[264]   P. H. HÜNENBERGER and M. REIF, *Single-Ion Solvation*, RSC Publishing, Cambridge UK, 2011.

[265]   Y. HYON, J. A. CARRILLO, Q. DU and C. LIU, *A Maximum Entropy Principle Based Closure Method for Macro-Micro Models of Polymeric Materials*, Kinetic and Related Models, 1 (2008), pp. 171-184.

[266]   Y. HYON, B. EISENBERG and C. LIU, *A Mathematical Model for the Hard Sphere Repulsion in Ionic Solutions*, Communications in Mathematical Sciences, 9 (2011), pp. 459–475  also available as preprint#  2318  (IMA,  University  of  Minnesota,  Minneapolis) http://www.ima.umn.edu/preprints/jun2010/jun2010.html, 2010.

[267]   Y. HYON, J. E. FONSECA, B. EISENBERG and C. LIU, *Energy variational approach to study charge inversion (layering) near charged walls*, Discrete and Continuous Dynamical Systems Series B (DCDS-B)  (2012, *in the press*).

[268]   Y. HYON, J. E. FONSECA, B. EISENBERG and C. LIU, *A New Poisson-Nernst-Planck Equation (PNP-FS-IF) for Charge Inversion Near Walls*, Biophysical Journal, 100 (2011), pp. 578a.

[269]   Y. HYON, D. Y. KWAK and C. LIU, *Energetic Variational Approach in Complex Fluids : Maximum Dissipation Principle*, Discrete and Continuous Dynamical Systems  (DCDS-A), 26 (2010), pp. 1291 - 1304, available at URL: http://www.ima.umn.edu as IMA Preprint Series # 2228.

[270]   J. G. IBARRA-ARMENTA, A. MARTIN-MOLINA and M. QUESADA-PEREZ, *Testing a modified model of the Poisson-Boltzmann theory that includes ion size effects through Monte Carlo simulations*, Physical Chemistry Chemical Physics, 11 (2009), pp. 309-316.

[271]   W. IM, M. S. LEE and C. L. BROOKS, 3RD, *Generalized born model with a simple smoothing function*, J Comput Chem, 24 (2003), pp. 1691-702.

[272]   W. IM and B. ROUX, *Brownian Dynamics simulations of ions channels: a general treatment of electrostatic reaction fields for molecular pores of arbitrary geometry*, Biophysical Journal, 115 (2001), pp. 4850-4861.

[273]   W. IM and B. ROUX, *Ion permeation and selectivity of OmpF porin: a theoretical study based on molecular dynamics, Brownian dynamics, and continuum electrodiffusion theory*, Journal of Molecular Biology, 322 (2002), pp. 851-69.

[274]   W. IM and B. ROUX, *Ions and counterions in a biological channel: a molecular dynamics simulation of OmpF porin from Escherichia coli in an explicit membrane with 1 M KCl aqueous salt solution*, Journal of Molecular Biology, 319 (2002), pp. 1177-97.

[275]   W. IM, S. SEEFELD and B. ROUX, *A Grand Canonical Monte Carlo-Brownian Dynamics Algorithm for Simulating Ion Channels*, Biophysical Journal, 79 (2000), pp. 788-801.

[276]   R. T. JACOBSEN, S. G. PENONCELLO, E. W. LEMMON and R. SPAN, *Multiparameter Equations of State*, in J. V. Sengers, R. F. Kayser, C. J. Peters and H. J. White, Jr., eds., *Equations of State for Fluids and Fluid Mixtures*, Elsevier, New York, 2000, pp. 849-882.

[277]   J. JANECEK and R. R. NETZ, *Effective screening length and quasiuniversality for the restricted primitive model of an electrolyte solution*, J Chem Phys, 130 (2009), pp. 074502-15.

[278]   D. J. JENDEN and A. S. FAIRHURST, *The pharmacology of ryanodine*, Pharmacol Rev, 21 (1969), pp. 1-25.

[279]   J. W. JEROME, *Analysis of Charge Transport. Mathematical Theory and Approximation of Semiconductor Models*, Springer-Verlag, New York, 1995.

[280]   D. JIMENEZ-MORALES, J. LIANG and B. EISENBERG, *Ionizable side chains at catalytic active sites of enzymes*, European Biophysics Journal, 41 (2012), pp. 449-460.

[281]   R. N. JORISSEN and M. K. GILSON, *Virtual screening of molecular databases using a support vector machine*, J Chem Inf Model, 45 (2005), pp. 549-61.

[282]   R. N. JORISSEN, G. S. REDDY, A. ALI, M. D. ALTMAN, S. CHELLAPPAN, S. G. ANJUM, B. TIDOR, C. A. SCHIFFER, T. M. RANA and M. K. GILSON, *Additivity in the analysis and design of HIV protease inhibitors*, J Med Chem, 52 (2009), pp. 737-54.

[283]   D. JOSEPH, *Fluid Dynamics of Viscoelastic Liquids*, Springer-Verlag, New York, 1990.

[284]   I. S. JOUNG and T. E. CHEATHAM, *Determination of Alkali and Halide Monovalent Ion Parameters for Use in Explicitly Solvated Biomolecular Simulations*, The Journal of Physical Chemistry B, 112 (2008), pp. 9020-9041.

[285]   P. JUNGWIRTH, B. J. FINLAYSON-PITTS and D. J. TOBIAS, *Introduction:  Structure and Chemistry at Aqueous Interfaces*, Chemical Reviews, 106 (2006), pp. 1137-1139.

[286]   P. JUNGWIRTH and B. WINTER, *Ions at Aqueous Interfaces: From Water Surface to Hydrated Proteins*, Annual Review of Physical Chemistry, 59 (2008), pp. 343-366.

[287]   J.-C. JUSTICE, *Conductance of Electrolyte Solutions*, in B. E. Conway, J. O. M. Bockris and E. Yaeger, eds., *Comprehensive Treatise of Electrochemistry Volume 5 Thermondynbamic and Transport Properties of Aqueous and Molten Electrolytes*, Plenum, New York, 1983, pp. 223-338.

[288]   V. KAIRYS, M. X. FERNANDES and M. K. GILSON, *Screening drug-like compounds by docking to homology models: a systematic study*, J Chem Inf Model, 46 (2006), pp. 365-79.

[289]   V. KAIRYS and M. K. GILSON, *Enhanced docking with the mining minima optimizer: acceleration and side-chain flexibility*, J Comput Chem, 23 (2002), pp. 1656-70.

[290]   V. KAIRYS, M. K. GILSON and M. X. FERNANDES, *Using protein homology models for structure-based studies: approaches to model refinement*, ScientificWorldJournal, 6 (2006), pp. 1542-54.

[291]   V. KAIRYS, M. K. GILSON, V. LATHER, C. A. SCHIFFER and M. X. FERNANDES, *Toward the design of mutation-resistant enzyme inhibitors: further evaluation of the substrate envelope hypothesis*, Chem Biol Drug Des, 74 (2009), pp. 234-45.

[292]   V. KAIRYS, M. K. GILSON and B. LUY, *Structural model for an AxxxG-mediated dimer of surfactant-associated protein C*, Eur J Biochem, 271 (2004), pp. 2086-92.

[293]   I. KALCHER and J. DZUBIELLA, *Structure-thermodynamics relation of electrolyte solutions*, J Chem Phys, 130 (2009), pp. 134507.

[294]   I. KALCHER, J. C. F. SCHULZ and J. DZUBIELLA, *Electrolytes in a nanometer slab-confinement: Ion-specific structure and solvation forces*, J Chem Phys, 133 (2010), pp. 164511-15.

[295]   I. KALCHER, J. C. F. SCHULZ and J. DZUBIELLA, *Ion-Specific Excluded-Volume Correlations and Solvation Forces*, Physical Review Letters, 104 (2010), pp. 097802.

[296]    Y. V. KALYUZHNYI, V. VLACHY and K. A. DILL, *Aqueous alkali halide solutions: can osmotic coefficients be explained on the basis of the ionic sizes alone?*, Physical Chemistry Chemical Physics, 12 (2010), pp. 6260-6266.

[297]    M. A. KASTENHOLZ and P. H. HUNENBERGER, *Computation of methodology-independent ionic solvation free energies from molecular simulations. I. The electrostatic potential in molecular liquids*, J Chem Phys, 124 (2006), pp. 124106-27.

[298]    M. A. KASTENHOLZ and P. H. HUNENBERGER, *Computation of methodology-independent ionic solvation free energies from molecular simulations. II. The hydration free energy of the sodium cation*, J Chem Phys, 124 (2006), pp. 224501-20.

[299]    M. A. KASTENHOLZ and P. H. HUNENBERGER, *Development of a lattice-sum method emulating nonperiodic boundary conditions for the treatment of electrostatic interactions in molecular simulations: A continuum-electrostatics study*, J Chem Phys, 124 (2006), pp. 124108-12.

[300]    I. V. KHAVRUTSKII, J. DZUBIELLA and J. A. MCCAMMON, *Computing accurate potentials of mean force in electrolyte solutions with the generalized gradient-augmented harmonic Fourier beads method*, J Chem Phys, 128 (2008), pp. 044106.

[301]    B. J. KILLIAN, J. Y. KRAVITZ, S. SOMANI, P. DASGUPTA, Y. P. PANG and M. K. GILSON, *Configurational entropy in protein-peptide binding: computational study of Tsg101 ubiquitin E2 variant domain with an HIV-derived PTAP nonapeptide*, J Mol Biol, 389 (2009), pp. 315-35.

[302]    B. J. KILLIAN, J. YUNDENFREUND KRAVITZ and M. K. GILSON, *Extraction of configurational entropy from molecular simulations via an expansion approximation*, J Chem Phys, 127 (2007), pp. 024107.

[303]    I. KLAPPER, R. HAGSTROM, R. FINE, K. SHARP and B. HONIG, *Focusing of electric fields in the active site of Cu-Zn superoxide dismutase: effects of ionic strength and amino-acid modification*, Proteins, 1 (1986), pp. 47-59.

[304]    M. M. KŁOSEK-DYGAS, B. M. HOFFMAN, B. J. MATKOWSKY, A. NITZAN, M. A. RATNER and Z. SCHUSS, *Diffusion theory of multidimensional activated rate processes: The role of anisotropy*, J Chem Phys, 90 (1989), pp. 1141-1148.

[305]    M. M. KŁOSEK, *Half-range expansion analysis for langevin dynamics in the high-friction limit with a singular absorbing boundary condition: Noncharacteristic case*, Journal of Statistical Physics, 79 (1995), pp. 313-345.

[306]    M. M. KŁOSEK, B. J. MATKOWSKY and Z. SCHUSS, *The Kramers problem in the turnover regime: The role of the stochastic separatrix.*, Berichte der Bunsen - Gesellschaft fur Physikalishe Chemie, 95 (1991), pp. 331-337.

[307]    H. KOKUBO, J. ROSGEN, D. W. BOLEN and B. M. PETTITT, *Molecular Basis of the Apparent Near Ideality of Urea Solutions*, Biophys. J., 93 (2007), pp. 3392-3407.

[308]    G. M. KONTOGEORGIS and G. K. FOLAS, *Thermodynamic Models for Industrial Applications: From Classical and Advanced Mixing Rules to Association Theories*, John Wiley & Sons, Ltd, 2009.

[309]    A. A. KORNYSHEV, *Double-Layer in Ionic Liquids: Paradigm Change?*, J. Phys. Chem. B, 111 (2007), pp. 5545-5557.

[310]    P. G. KOSTYUK, O. A. KRISHTAL and P. A. DOROSHENKO, *Calcium currents in snail neurones. I. Identification of calcium current*, Pflugers Arch, 348 (1974), pp. 83-93.

[311]    H. A. KRAMERS, *Brownian motion in a field of force and the diffusion model of chemical reactions.*, Physica, 7 (1940), pp. 284–304.

[312]    C. A. KRAUS, *The present status of the theory of electrolytes*, Bull. Amer. Math. Soc., 44 (1938), pp. 361-383.

[313]    D. KRAUSS, B. EISENBERG and D. GILLESPIE, *Selectivity sequences in a model calcium channel: role of electrostatic field strength*, European Biophysics Journal, 40 (2011), pp. 775-782.

[314]    D. KRAUSS and D. GILLESPIE, *Sieving experiments and pore diameter: It's not a simple relationship*, European Biophysics Journal, 39 (2010), pp. 1513-1521.

[315]    I. KRON, S. MARSHALL, P. MAY, G. HEFTER and E. KÖNIGSBERGER, *The ionic product of water in highly concentrated aqueous electrolyte solutions*, Monatshefte für Chemie / Chemical Monthly, 126 (1995), pp. 819-837.

[316]    A. KUMAR and V. S. PATWARDHAN, *Activity coefficients in mixed aqueous electrolyte solutions with a common ion*, AIChE Journal, 38 (1992), pp. 793-796.

[317]    P. V. KUMAR and M. MARONCELLI, *The non-separability of ``dielectric'' and ``mechanical'' friction in molecular systems: A simulation study*, J Chem Phys, 112 (2000), pp. 5370-5381.

[318]    W. KUNZ, *Specific Ion Effects*, World Scientific Singapore, 2009.

[319]    W. KUNZ and R. NEUEDER, *An Attempt at an Overview*, in W. Kunz, ed., *Specific Ion Effects*, World Scientific Singapore, 2009, pp. 11-54.

[320]    M. G. KURNIKOVA, R. D. COALSON, P. GRAF and A. NITZAN, *A Lattice Relaxation Algorithm for 3D Poisson-Nernst-Planck Theory with Application to Ion Transport Through the Gramicidin A Channel*, Biophysical Journal, 76 (1999), pp. 642-656.

[321]    K. J. LAIDLER, J. H. MEISER and B. C. SANCTUARY, *Physical Chemistry*, BrooksCole, Belmont CA, 2003.

[322]    R. G. LARSON, *The Structure and Rheology of Complex Fluids* Oxford, New York, 1995.

[323]    L. L. LEE, *Molecular Thermodynamics of Electrolyte Solutions*, World Scientific Singapore, 2008.

[324]    L. L. LEE, *Molecular Thermodynamics of Nonideal Fluids* Butterworth-Heinemann, New York, 1988.

[325]    D. LEVITT, *General Continuum theory for a multiion channel*, Biophysical Journal, 59 (1991), pp. 271-277.

[326]    D. LEVITT, *General Continuum theory for a multiion channel. Application for a multiion channel*, Biophysical Journal, 59 (1991), pp. 278-288.

[327]    D. G. LEVITT, *Comparison of Nernst-Planck and reaction-rate models for multiply occupied channels.*, Biophys. J, 37 (1982), pp. 575–587.

[328]    D. G. LEVITT, *Electrostatic calculations for an ion channel. I. Energy and potential profiles and interactions between ions.*, Biophys. J., 22 (1978), pp. 209-219.

[329]    D. G. LEVITT, *Electrostatic calculations for an ion channel. II. Kinetic behavior of the gramicidin A channel*, Biophys J, 22 (1978), pp. 221-48.

[330]    D. G. LEVITT, *General continuum analysis of transport through pores. I. Proof of Onsager's reciprocity postulate for uniform pore*, Biophys J, 15 (1975), pp. 533-51.

[331]    D. G. LEVITT, *General continuum analysis of transport through pores. II. Nonuniform pores*, Biophys J, 15 (1975), pp. 553-63.

[332]    D. G. LEVITT, *Interpretation of biological ion channel flux data. Reaction rate versus continuum theory*, Ann. Rev. Biophys. Biophys. Chem, 15 (1986), pp. 29–57.

[333]    D. G. LEVITT, *Modeling of ion channels*, J Gen Physiol, 113 (1999), pp. 789-94.

[334]    D. G. LEVITT, *Strong electrolyte continuum theory solution for equilibrium profiles, diffusion limitation, and conductance in charged ion channels.*, Biophys. J., 52 (1985), pp. 575–587.

[335]    B. LI, *Continuum electrostatics for ionic solutions with non-uniform ionic sizes*, Nonlinearity, 22 (2009), pp. 811.

[336]   B. LI, *Minimization of Electrostatic Free Energy and the Poisson--Boltzmann Equation for Molecular Solvation with Implicit Solvent*, SIAM Journal on Mathematical Analysis, 40 (2009), pp. 2536-2566.

[337]   B. LI, X. CHENG and Z. ZHANG, *Dielectric Boundary Force in Molecular Solvation with the Poisson--Boltzmann Free Energy: A Shape Derivative Approach*, SIAM Journal on Applied Mathematics, 71 (2011), pp. 2093-2111.

[338]   S. C. LI, M. HOYLES, S. KUYUCAK and S. H. CHUNG, *Brownian dynamics study of ion transport in the vestibule of membrane channels*, Biophys J, 74 (1998), pp. 37-47.

[339]   Y. A. LI, P. J. OLVER and P. ROSENAU, *Non-analytic solutions of nonlinear wave models*, IMA Preprint Series Institute for Mathematics and its Applications University of Minnesota, # 1591 (1998), pp. 15.

[340]   B. LIN, K. Y. WONG, C. HU, H. KOKUBO and B. M. PETTITT, *Fast Calculations of Electrostatic Solvation Free Energy from Reconstructed Solvent Density using proximal Radial Distribution Functions*, The Journal of Physical Chemistry Letters, 2 (2011), pp. 1626-1632.

[341]   F.-H. LIN, C. LIU and P. ZHANG, *On a Micro-Macro Model for Polymeric Fluids near Equilibrium*, Communications on Pure and Applied Mathematics, 60 (2007), pp. 838-866.

[342]   F.-H. LIN, C. LIU and P. ZHANG, *On hydrodynamics of viscoelastic fluids*, Communications on Pure and Applied Mathematics, 58 (2005), pp. 1437-1471.

[343]   Y. LIN, K. THOMEN and J.-C. D. HEMPTINNE, *Multicomponent Equations of State for Electrolytes*, American Institute of Chemical Engineers AICHE Journal, 53 (2007), pp. 989-1005.

[344]   F. LIPPARINI, G. SCALMANI, B. MENNUCCI, E. CANCES, M. CARICATO and M. J. FRISCH, *A variational formulation of the polarizable continuum model*, J Chem Phys, 133 (2010), pp. 014106-11.

[345]   C. LIU, *An Introduction of Elastic Complex Fluids: An Energetic Variational Approach*, in T. Y. Hou, Liu, C., Liu, J.-g, ed., *Multi-scale Phenomena in Complex Fluids: Modeling, Analysis and Numerical Simulations*, World Scientific Publishing Company, Singapore, 2009.

[346]   T. LIU, Y. LIN, X. WEN, R. N. JORISSEN and M. K. GILSON, *BindingDB: a web-accessible database of experimentally determined protein-ligand binding affinities*, Nucleic Acids Res, 35 (2007), pp. D198-201.

[347]   W. LIU and B. WANG, *Poisson-Nernst-Planck systems for narrow tubular-like membrane channels,*, J. Dynam. Differential Equations, 22 (2010), pp. 413-437.

[348]   J. R. LOEHE and M. D. DONOHUE, *Recent advances in modeling thermodynamic properties of aqueous strong electrolyte systems*, AIChE Journal, 43 (1997), pp. 180-195.

[349]   B. LU, D. ZHANG and J. A. MCCAMMON, *Computation of electrostatic forces between solvated molecules determined by the Poisson--Boltzmann equation using a boundary element method*, J Chem Phys, 122 (2005), pp. 214102-7.

[350]   D. G. LUCHINSKY, R. TINDJONG, I. KAUFMAN, P. V. E. MCCLINTOCK and R. S. EISENBERG, *Self-consistent analytic solution for the current and the access resistance in open ion channels*, Physical Review E (Statistical, Nonlinear, and Soft Matter Physics), 80 (2009), pp. 021925-12.

[351]   M. LUNDSTROM, *APPLIED PHYSICS: Enhanced: Moore's Law Forever?*, Science, 299 (2003), pp. 210-211.

[352]   M. LUNDSTROM, *Fundamentals of Carrier Transport*, Addison-Wesley, NY, 2000.

[353]   R. LUO, L. DAVID and M. K. GILSON, *Accelerated Poisson-Boltzmann calculations for static and dynamic systems*, J Comput Chem, 23 (2002), pp. 1244-53.

[354]   R. LUO, H. S. GILSON, M. J. POTTER and M. K. GILSON, *The physical basis of nucleic acid base stacking in water*, Biophys J, 80 (2001), pp. 140-8.

[355]   R. LUO, M. S. HEAD, J. A. GIVEN and M. K. GILSON, *Nucleic acid base-pairing and N-methylacetamide self-association in chloroform: affinity and conformation*, Biophys Chem, 78 (1999), pp. 183-93.

[356]   Y. LUO and B. T. ROUX, *Simulation of Osmotic Pressure in Concentrated Aqueous Salt Solutions*, The Journal of Physical Chemistry Letters, 1 (2009), pp. 183-189.

[357]   J. R. MACDONALD, *Theory of ac Space-Charge Polarization Effects in Photoconductors, Semiconductors, and Electrolytes*, Physical Review, 92 (1953), pp. 4-17.

[358]   J. R. MACDONALD and D. R. FRANCESCHETTI, *Theory of small-signal ac response of solids and liquids with recombining mobile charge*, J Chem Phys, 68 (1978), pp. 1614-1637.

[359]   R. MACKINNON, *Nobel Lecture. Potassium channels and the atomic basis of selective ion conduction*, Biosci Rep, 24 (2004), pp. 75-100.

[360]   A. B. MAMONOV, R. D. COALSON, A. NITZAN and M. G. KURNIKOVA, *The role of the dielectric barrier in narrow biological channels: a novel composite approach to modeling single-channel currents*, Biophys J, 84 (2003), pp. 3646-61.

[361]   A. B. MAMONOV, M. G. KURNIKOVA and R. D. COALSON, *Diffusion constant of K+ inside Gramicidin A: a comparative study of four computational methods*, Biophys Chem, 124 (2006), pp. 268-78.

[362]   K. L. MARDIS, R. LUO and M. K. GILSON, *Interpreting trends in the binding of cyclic ureas to HIV-1 protease*, J Mol Biol, 309 (2001), pp. 507-17.

[363]   P. A. MARKOWICH, C. A. RINGHOFER and C. SCHMEISER, *Semiconductor Equations*, Springer-Verlag, New York, 1990.

[364]   D. MARREIRO, *Efficient partile-based simulation of ion channels*, Electrical Engineering, Illinois Institute of Technology (IIT), Chicago, 2006, pp. 146.

[365]   E. W. MCCLESKEY, *Ion channel selectivity using an electric stew*, Biophys J, 79 (2000), pp. 1691-2.

[366]   E. W. MCCLESKEY and W. ALMERS, *The Ca channel in skeletal muscle is a large pore*, Proceedings of the National Academy of Science USA, 82 (1985), pp. 7149-7153.

[367]   E. W. MCCLESKEY, M. D. WOMACK and L. A. FIEBER, *Structural properties of voltage-dependent calcium channels*, Int Rev Cytol, 137C (1993), pp. 39-54.

[368]   A. MCNABB and L. BASS, *Flux-ratio theorems for nonlinear equations of generalized diffusion*, IMA Journal Applied Mathematics, 43 (1989), pp. 1-9.

[369]   A. MCNABB and L. BASS, *Flux theorems for linear multicomponent diffusion*, IMA Journal Applied Mathematics, 43 (1990), pp. 155-161.

[370]   G. MEISSNER, *Regulation of mammalian ryanodine receptors*, Front Biosci, 7 (2002), pp. d2072-80.

[371]   G. MEISSNER, *Ryanodine activation and inhibition of the $Ca^{2+}$ release channel of sarcoplasmic reticulum.*, J Biol Chem, 261 (1986), pp. 6300-6306.

[372]   H. MIEDEMA, A. METER-ARKEMA, J. WIERENGA, J. TANG, B. EISENBERG, W. NONNER, H. HEKTOR, D. GILLESPIE and W. MEIJBERG, *Permeation properties of an engineered bacterial OmpF porin containing the EEEE-locus of Ca2+ channels*, Biophys J, 87 (2004), pp. 3137-47.

[373]   H. MIEDEMA, M. VROUENRAETS, J. WIERENGA, D. GILLESPIE, B. EISENBERG, W. MEIJBERG and W. NONNER, *Ca2+ selectivity of a chemically modified OmpF with reduced pore volume*, Biophys J, 91 (2006), pp. 4392-400.

[374]   M. MIHAILESCU and M. K. GILSON, *On the theory of noncovalent binding*, Biophys J, 87 (2004), pp. 23-36.

[375]   S. MOGHADDAM, Y. INOUE and M. K. GILSON, *Host-guest complexes with protein-ligand-like affinities: computational analysis and design*, J Am Chem Soc, 131 (2009), pp. 4012-21.

[376]    S. MOGHADDAM, C. YANG, M. REKHARSKY, Y. H. KO, K. KIM, Y. INOUE and M. K. GILSON, *New ultrahigh affinity host-guest complexes of cucurbit[7]uril with bicyclo[2.2.2]octane and adamantane guests: thermodynamic analysis and evaluation of M2 affinity calculations*, J Am Chem Soc, 133 (2011), pp. 3570-81.

[377]    J. J. MOLINA, J. F. DUFRE CHE, M. SALANNE, O. BERNARD and P. TURQ, *Primitive models of ions in solution from molecular descriptions: A perturbation approach*, J Chem Phys, 135 (2011), pp. 234509.

[378]    J. J. MOLINA, J.-F. DUFRECHE, M. SALANNE, O. BERNARD and P. TURQ, *Primitive models of ions in solution from molecular descriptions: A perturbation approach*, J Chem Phys, 135 (2011), pp. 234509-20.

[379]    J. MONGAN, D. A. CASE and J. A. MCCAMMON, *Constant pH molecular dynamics in generalized Born implicit solvent*, J Comput Chem, 25 (2004), pp. 2038-48.

[380]    G. E. MOORE, *Cramming more components onto integrated circuits*, Electronics Magazine., 38 (1965).

[381]    G. E. MOORE, *Lithography and the future of Moore's law*, in R. D. Allen, ed., SPIE, Santa Clara, CA, USA, 1995, pp. 2-17.

[382]    Y. MORI, C. LIU and R. S. EISENBERG, *A model of electrodiffusion and osmotic water flow and its energetic structure*, Physica D: Nonlinear Phonomena 240 (2011), pp. 1835-1852.

[383]    G. MOY, B. CORRY, S. KUYUCAK and S. H. CHUNG, *Tests of continuum theories as models of ion channels. I. Poisson-Boltzmann theory versus Brownian dynamics*, Biophys J, 78 (2000), pp. 2349-63.

[384]    J. A. MYERS, S. I. SANDLER and R. H. WOOD, *An Equation of State for Electrolyete Solutions Covering Wide Ranges of Temperature, Pressure, and Composition*, Industrial and Engineering Chemical Research, 41 (2002).

[385]    B. NADLER, U. HOLLERBACH and R. S. EISENBERG, *Dielectric boundary force and its crucial role in gramicidin*, Phys Rev E Stat Nonlin Soft Matter Phys, 68 (2003), pp. 021905.

[386]    B. NADLER, Z. SCHUSS and A. SINGER, *Langevin Trajectories between fixed concentrations*, Physical Review Letters., in the press (2005).

[387]    B. NADLER, Z. SCHUSS, A. SINGER and B. EISENBERG, *Diffusion through protein channels: from molecular description to continuum equations.*, Nanotechnology, 3 (2003), pp. 439.

[388]    T. NAEH, M. M. KŁOSEK, B. J. MATKOWSKY and Z. SCHUSS, *A direct approach to the exit problem.*, Siam J. Appl. Math, 50 (1990), pp. 595-627.

[389]    T. NAGRANI, M. SIYAMWALA, G. VAHID and S. BEKHEIT, *Ryanodine calcium channel: a novel channelopathy for seizures*, The neurologist, 17 (2011), pp. 91-4.

[390]    M. N. NALAM, A. ALI, M. D. ALTMAN, G. S. REDDY, S. CHELLAPPAN, V. KAIRYS, A. OZEN, H. CAO, M. K. GILSON, B. TIDOR, T. M. RANA and C. A. SCHIFFER, *Evaluating the substrate-envelope hypothesis: structural analysis of novel HIV-1 protease inhibitors designed to be robust against drug resistance*, J Virol, 84 (2010), pp. 5368-78.

[391]    J. NEWMAN and K. E. THOMAS-ALYEA, *Electrochemical Systems*, Wiley-Interscience, New York, 2004.

[392]    J. E. NIELSEN and J. A. MCCAMMON, *Calculating pKa values in enzyme active sites*, Protein Sci, 12 (2003), pp. 1894-901.

[393]    A. NITZAN and Z. SCHUSS, *Multidimensional Barrier Crossing*, in G. Fleming and P. Hänggi, eds., *Activated Barrier Crossing: Applications in Physics, Chemistry and Biology*, World Scientific Publishing, New Jersey, 1993, pp. 42-81.

[394]  W. NONNER, D. P. CHEN and B. EISENBERG, *Progress and Prospects in Permeation*, Journal of General Physiology, 113 (1999), pp. 773-782.

[395]  W. NONNER and B. EISENBERG, *Ion Permeation and Glutamate Residues Linked by Poisson-Nernst-Planck Theory in L-type Calcium Channels*, Biophys. J., 75 (1998), pp. 1287-1305.

[396]  W. NONNER, D. GILLESPIE, D. HENDERSON and B. EISENBERG, *Ion accumulation in a biological calcium channel: effects of solvent and confining pressure*, J Physical Chemistry B, 105 (2001), pp. 6427-6436.

[397]  W. W. PARSON, Z. T. CHU and A. WARSHEL, *Electrostatic control of charge separation in bacterial photosynthesis*, Biochim Biophys Acta, 1017 (1990), pp. 251-72.

[398]  V. S. PATWARDHAN and A. KUMAR, *Thermodynamic properties of aqueous solutions of mixed electrolytes: A new mixing rule*, AIChE Journal, 39 (1993), pp. 711-714.

[399]  V. S. PATWARDHAN and A. KUMAR, *A unified approach for prediction of thermodynamic properties of aqueous mixed-electrolyte solutions. Part II: Volume, thermal, and other properties*, AIChE Journal, 32 (1986), pp. 1429-1438.

[400]  H. PAUL and S. MATTHIAS, *Binary non-additive hard sphere mixtures: fluid demixing, asymptotic decay of correlations and free fluid interfaces*, Journal of Physics: Condensed Matter, 22 (2010), pp. 325108.

[401]  J. PAYANDEH, T. SCHEUER, N. ZHENG and W. A. CATTERALL, *The crystal structure of a voltage-gated sodium channel*, Nature, 475 (2011), pp. 353-358.

[402]  H. PEARSON, *Protein engineering: The fate of fingers*, Nature, 455 (2008), pp. 160-164.

[403]  P. B. PETERSEN and R. J. SAYKALLY, *On the Nature of Ions at the Liquid Water Surface*, Annual Review of Physical Chemistry, 57 (2006), pp. 333-364.

[404]  J. PICALEK, B. MINOFAR, J. KOLAFA and P. JUNGWIRTH, *Aqueous solutions of ionic liquids: study of the solution/vapor interface using molecular dynamics simulations*, Physical chemistry chemical physics : PCCP, 10 (2008), pp. 5765-75.

[405]  R. F. PIERRET, *Semiconductor Device Fundamentals*, Addison Wesley, New York, 1996.

[406]  K. S. PITZER, *Activity Coefficients in Electrolyte Solutions*, CRC Press, Boca Raton FL USA, 1991.

[407]  K. S. PITZER, *Thermodynamics*, McGraw Hill, New York, 1995.

[408]  K. S. PITZER and J. J. KIM, *Thermodynamics of electrolytes. IV. Activity and osmotic coefficients for mixed electrolytes*, Journal of the American Chemical Society, 96 (1974), pp. 5701-5707.

[409]  E. POLLAK, A. BEREZHKOVSKII and Z. SCHUSS, *Activated rate processes: a relation between Hamiltonian and stochastic theories*, J Chem Phys, 100 (1994), pp. 334-339.

[410]  R. M. PYTKOWICZ, *Activity Coefficients in Electrolyte Solutions*, CRC, Boca Raton FL USA, 1979.

[411]  A. A. RASHIN and B. HONIG, *Reevaluation of the Born model of ion hydration*, J Phys Chem B, 89 (1985), pp. 5588-5593.

[412]  M. M. REIF and P. H. HUNENBERGER, *Computation of methodology-independent single-ion solvation properties from molecular simulations. III. Correction terms for the solvation free energies, enthalpies, entropies, heat capacities, volumes, compressibilities, and expansivities of solvated ions*, J Chem Phys, 134 (2011), pp. 144103-30.

[413]  M. M. REIF and P. H. HUNENBERGER, *Computation of methodology-independent single-ion solvation properties from molecular simulations. IV. Optimized Lennard-Jones interaction parameter sets for the alkali and halide ions in water*, J Chem Phys, 134 (2011), pp. 144104-25.

[414]  W. H. REINMUTH, *Theory of Stationary Electrode Polarography*, Anal Chem, 33 (1961), pp. 1793-1794.

[415]  M. V. REKHARSKY, T. MORI, C. YANG, Y. H. KO, N. SELVAPALAM, H. KIM, D. SOBRANSINGH, A. E. KAIFER, S. LIU, L. ISAACS, W. CHEN, S. MOGHADDAM, M. K. GILSON, K. KIM and Y. INOUE, *A*

synthetic host-guest system achieves avidin-biotin affinity by overcoming enthalpy-entropy compensation, Proc Natl Acad Sci U S A, 104 (2007), pp. 20737-42.

[416]   P. M. V. RESIBOIS, *Electrolyte Theory*, Harper & Row, New York, 1968.

[417]   S. A. RICE and P. GRAY, *Statistical Mechanics of Simple Fluids*, Interscience (Wiley), New York, 1965.

[418]   R. R. RICHARDS, *Chemical potential and osmotic pressure*, Journal of Chemical Education, 56 (1979), pp. 579-null.

[419]   M. RIORDAN and L. HODDESON, *Crystal Fire*, Norton, New York, 1997.

[420]   R. A. ROBINSON and R. H. STOKES, *Electrolyte Solutions*, Butterworths Scientific Publications, also Dover books, 2002., London, 1959.

[421]   G. M. ROGER, S. DURAND-VIDAL, O. BERNARD and P. TURQ, *Electrical conductivity of mixed electrolytes: Modeling within the mean spherical approximation*, The journal of physical chemistry. B, 113 (2009), pp. 8670-6.

[422]   Y. ROSENFELD, *Equation of state and correlation functions of strongly coupled plasma mixtures: Density functional theory and analytic models*, Physical Review. E. Statistical Physics, Plasmas, Fluids, and Related Interdisciplinary Topics, 54 (1996), pp. 2827-2838.

[423]   Y. ROSENFELD, *Free-energy model for charged Yukawa mixtures: Asymptotic strong-coupling limit and a nonlinear mixing rule*, Physical Review. E. Statistical Physics, Plasmas, Fluids, and Related Interdisciplinary Topics, 47 (1993), pp. 2676-2682.

[424]   J. ROSGEN, B. M. PETTITT, J. PERKYNS and D. W. BOLEN, *Statistical Thermodynamic Approach to the Chemical Activities in Two-Component Solutions*, J. Phys. Chem. B, 108 (2004), pp. 2048-2055.

[425]   R. ROTH, *Fundamental measure theory for hard-sphere mixtures: a review*, Journal of Physics: Condensed Matter, 22 (2010), pp. 063102.

[426]   R. ROTH, R. EVANS, A. LANG and G. KAHL, *Fundamental measure theory for hard-sphere mixtures revisited: the White Bear version*, J. Phys.: Condens. Matter 14 (2002), pp. 12063-12078.

[427]   D. J. ROUSTON, *Bipolar Semiconductor Devices*, McGraw-Hill Publishing Company,, New York, 1990.

[428]   B. ROUX, *Implicit solvent models*, in O. Becker, A. D. MacKerrel, R. B. and M. Watanabe, eds., *Computational Biophysics*, Marcel Dekker Inc, New York, 2001, pp. p. 133-155.

[429]   J. S. ROWLINSON, *The Perfect Gas*, Macmillan, New York, 1963.

[430]   I. RUBINSTEIN, *Electro-diffusion of ions*, SIAM, Philadelphia, 1990.

[431]   S. T. RUSSELL and A. WARSHEL, *Calculations of electrostatic energies in proteins. The energetics of ionized groups in bovine pancreatic trypsin inhibitor*, J Mol Biol, 185 (1985), pp. 389-404.

[432]   G. B. RUTKAI, D. BODA and T. S. KRISTÓF, *Relating Binding Affinity to Dynamical Selectivity from Dynamic Monte Carlo Simulations of a Model Calcium Channel*, The Journal of Physical Chemistry Letters, 1 (2010), pp. 2179-2184.

[433]   R. RYHAM, F. COHEN and R. S. EISENBERG, *A Dynamic Model of Open Vesicles in Fluids*, Communications in Mathematical Sciences, *(in the press)* (2012).

[434]   B. SABASS and U. SEIFERT, *Nonlinear, electrocatalytic swimming in the presence of salt*, J Chem Phys, 136 (2012), pp. 214507-13.

[435]   B. SAKMANN and E. NEHER, *Single Channel Recording.*, Plenum, New York, 1995.

[436]   J. SALA, E. GUARDIA and J. MARTI, *Effects of concentration on structure, dielectric, and dynamic properties of aqueous NaCl solutions using a polarizable model*, J Chem Phys, 132 (2010), pp. 214505-11.

[437]   M. SALANNE, C. SIMON, P. TURQ and P. A. MADDEN, *Conductivity-viscosity-structure: unpicking the relationship in an ionic liquid*, The journal of physical chemistry. B, 111 (2007), pp. 4678-84.

[438]   B. SAMANTHA, M. MARCO and C. CINZIA, *Solvation thermodynamics of alkali and halide ions in ionic liquids through integral equations*, The Journal of Chemical Physics, 129 (2008), pp. 074509.

[439]   M. SARANITI, S. ABOUD and R. EISENBERG, *The Simulation of Ionic Charge Transport in Biological Ion Channels: an Introduction to Numerical Methods*, Reviews in Computational Chemistry, 22 (2006), pp. 229-294.

[440]   W. A. SATHER and E. W. MCCLESKEY, *Permeation and selectivity in calcium channels*, Annu Rev Physiol, 65 (2003), pp. 133-59.

[441]   D. T. SAWYER, A. SOBKOWIAK and J. L. ROBERTS, JR., *Electrochemistry for Chemists*, 1995.

[442]   W. SCHMICKLER, *Interfacial Electrochemistry*, Oxford University Press, NY, 1996.

[443]   Z. SCHUSS, *Equilibrium and Recrossing of the Transition State: what can be learned from diffusion?*, Unpublished Manuscript, available by anonymous ftp from ftp.rush.edu in /pub/Eisenberg/Schuss/Rate Theory (1989).

[444]   Z. SCHUSS, *Singular perturbation methods for stochastic differential equations of mathematical physics.*, SIAM Review, 22 (1980), pp. 116-155.

[445]   Z. SCHUSS, *Theory and Applications of Stochastic Differential Equations*, John Wiley New York, 1980.

[446]   Z. SCHUSS, *Theory And Applications Of Stochastic Processes: An Analytical Approach* Springer, New York, 2009.

[447]   Z. SCHUSS, B. NADLER and R. S. EISENBERG, *Derivation of PNP Equations in Bath and Channel from a Molecular Model*, Physical Review E, 64 (2001), pp. 036116 1-14.

[448]   Z. SCHUSS, B. NADLER and R. S. EISENBERG, *Derivation of Poisson and Nernst-Planck equations in a bath and channel from a molecular model*, Phys Rev E Stat Nonlin Soft Matter Phys, 64 (2001), pp. 036116.

[449]   Z. SCHUSS, B. NADLER, A. SINGER and R. EISENBERG, *A PDE formulation of non-equilibrium statistical mechanics for ionic permeation,*, in S. M. Bezrukov, ed., *AIP Conference Proceedings , 3-6 September 2002: Unsolved Problems Of Noise And Fluctuations, UPoN 2002, 3rd International Conference on Unsolved Problems of Noise and Fluctuations in Physics, Biology, and High Technology* AIP, Washington, DC,, 2002.

[450]   C. N. SCHUTZ and A. WARSHEL, *What are the dielectric "constants" of proteins and how to validate electrostatic models?*, Proteins, 44 (2001), pp. 400-17.

[451]   S. SELBERHERR, *Analysis and Simulation of Semiconductor Devices*, Springer-Verlag, New York, 1984.

[452]   J. V. SENGERS, R. F. KAYSER, C. J. PETERS and H. J. WHITE, JR., *Equations of State for Fluids and Fluid Mixtures (Experimental Thermodynamics)* Elsevier, New York, 2000.

[453]   K. SHARP, A. JEAN-CHARLES and B. HONIG, *A local dielectric constant model for solvation free energies which accounts for solute polarizability*, J Phys Chem B, 96 (1992), pp. 3822-3828.

[454]   K. A. SHARP and B. HONIG, *Calculating total electrostatic energies with the nonlinear Poisson-Boltzmann equation*, J Phys Chem B, 94 (1990), pp. 7684-7692.

[455]   K. A. SHARP and B. HONIG, *Electrostatic Interactions in Macromolecules: Theory and Applications*, Annu Rev Biophys Biophys Chem, 19 (1990), pp. 301-332.

[456]   P. SHENG, J. ZHANG and C. LIU, *Onsager Principle and Electrorheological Fluid Dynamics*, Progress of Theoretical Physics Supplement No. 175 (2008), pp. 131-143.

[457]    W. SHOCKLEY, *Electrons and Holes in Semiconductors to applications in transistor electronics*, van Nostrand, New York, 1950.

[458]    M. SHUR, *Physics of Semiconductor Devices*, Prentice Hall, New York, 1990.

[459]    A. SHURKI, M. STRAJBL, C. N. SCHUTZ and A. WARSHEL, *Electrostatic basis for bioenergetics*, Methods Enzymol, 380 (2004), pp. 52-84.

[460]    A. SHURKI, M. STRAJBL, J. VILLA and A. WARSHEL, *How much do enzymes really gain by restraining their reacting fragments?*, J Am Chem Soc, 124 (2002), pp. 4097-107.

[461]    J. N. SHURKIN, *Broken Genius: The Rise and Fall of William Shockley, Creator of the Electronic Age*, Macmillan, New York, 2006.

[462]    J.-P. SIMONIN, L. BLUM and P. TURQ, *Real Ionic Solutions in the Mean Spherical Approximation. 1. Simple Salts in the Primitive Model*, Journal of Physical Chemistry, 100 (1996), pp. 7704-7709.

[463]    A. SINGER, D. GILLESPIE, J. NORBURY and R. S. EISENBERG, *Singular perturbation analysis of the steady-state Poisson–Nernst–Planck system: Applications to ion channels*, European Journal of Applied Mathematics, 19 (2008), pp. 541-560.

[464]    A. SINGER and J. NORBURY, *A Poisson-Nernst-Planck Model for Biological Ion Channels---An Asymptotic Analysis in a Tthree-dimensional Narrow Funnel*, SIAM J Appl Math, 70 (2009), pp. 949-968.

[465]    A. SINGER and Z. SCHUSS, *Brownian simulations and uni-directional flux in diffusion*, Phys. Rev. E, 71 (2005), pp. 026115.

[466]    S. SOMANI and M. K. GILSON, *Accelerated convergence of molecular free energy via superposition approximation-based reference states*, J Chem Phys, 134 (2011), pp. 134107.

[467]    S. SOMANI, B. J. KILLIAN and M. K. GILSON, *Sampling conformations in high dimensions using low-dimensional distribution functions*, J Chem Phys, 130 (2009), pp. 134102.

[468]    G. STELL and C. G. JOSLIN, *The Donnan Equilibrium: A Theoretical Study of the Effects of Interionic Forces*, Biophys J, 50 (1986), pp. 855-859.

[469]    B. G. STREETMAN, *Solid State Electronic Devices*, Prentice Hall, Englewood Cliffs, NJ, 1972.

[470]    L. STRYER, *Biochemistry*, W.H. Freeman, New York, 1995.

[471]    J. M. SWANSON, J. MONGAN and J. A. MCCAMMON, *Limitations of atom-centered dielectric functions in implicit solvent models*, J Phys Chem B Condens Matter Mater Surf Interfaces Biophys, 109 (2005), pp. 14769-72.

[472]    S. M. SZE, *Physics of Semiconductor Devices*, John Wiley & Sons, New York, 1981.

[473]    C. TANFORD, *Physical Chemistry of Macromolecuiles*, Wiley, New York, 1961.

[474]    C. TANFORD and J. REYNOLDS, *Nature's Robots: A History of Proteins*, Oxford, New York, 2001.

[475]    R. TINDJONG, I. KAUFMAN, P. V. E. MCCLINTOCK, D. G. LUCHINSKY and R. S. EISENBERG, *Nonequilibrium rate theory for conduction in open ion channels*, Fluctuation and Noise Letters, 11 (2012), pp. 83-93.

[476]    J. TOMASI, B. MENNUCCI and R. CAMMI, *Quantum Mechanical Continuum Solvation Models*, Chemical Reviews 105 (2005), pp. 2999-3093.

[477]    G. M. TORRIE and A. VALLEAU, *Electrical Double Layers: 4. Limitations of the Gouy-Chapman Theory*, Journal of Physical Chemistry, 86 (1982), pp. 3251-3257.

[478]    D. TOSTESON, *Membrane Transport: People and Ideas*, American Physiological Society, Bethesda MD, 1989.

[479]    J. TRYLSKA, J. ANTOSIEWICZ, M. GELLER, C. N. HODGE, R. M. KLABE, M. S. HEAD and M. K. GILSON, *Thermodynamic linkage between the binding of protons and inhibitors to HIV-1 protease*, Protein Sci, 8 (1999), pp. 180-95.

[480]   A. VAINRUB and B. M. PETTITT, *Accurate prediction of binding thermodynamics for DNA on surfaces*, The journal of physical chemistry. B, 115 (2011), pp. 13300-3.

[481]   T. A. VAN DER STRAATEN, R. S. EISENBERG, J. M. TANG, U. RAVAIOLI and N. ALURU, *Three dimensional Poisson Nernst Planck Simulation of ompF porin.*, Biophysical Journal, 80 (2001), pp. 115a.

[482]   T. A. VAN DER STRAATEN, J. TANG, R. S. EISENBERG, U. RAVAIOLI and N. R. ALURU, *Three-dimensional continuum simulations of ion transport through biological ion channels: effects of charge distribution in the constriction region of porin.* , J. Computational Electronics 1(2002), pp. 335-340.

[483]   T. A. VAN DER STRAATEN, J. M. TANG, R. S. EISENBERG, U. RAVAIOLI, N. ALURU, S. VARMA and J. E., *A study of mutations of ompf porin using Poisson-Nernst-Planck theory.* , Biophys. J., 82 (2002), pp. 207a.

[484]   T. A. VAN DER STRAATEN, J. M. TANG, U. RAVAIOLI, R. S. EISENBERG and N. R. ALURU, *Simulating Ion Permeation Through the OmpF Porin Ion channel Using Three-Dimensional Drift-Diffusion Theory*, Journal of Computational Electronics, 2 (2003), pp. 29-47.

[485]   N. G. VAN KAMPEN, *Stochastic Processes in Physics and Chemistry*, North Holland, New York, 1981.

[486]   W. VAN ROOSBROECK, *Theory of flow of electrons and holes in germanium and other semiconductors*, Bell System Technical Journal, 29 (1950), pp. 560-607.

[487]   K. VARSOS, J. LUNTZ, M. WELSH and K. SARABANDI, *Electric Field-Shaping Microdevices for Manipulation of Collections of Microscale Objects*, Proceedings of the IEEE, 99 (2011), pp. 2112-2124.

[488]   D. VASILESKA, S. M. GOODNICK and G. KLIMECK, *Computational Electronics: Semiclassical and Quantum Device Modeling and Simulation*, CRC Press, New York, 2010.

[489]   J. VILLA, M. STRAJBL, T. M. GLENNON, Y. Y. SHAM, Z. T. CHU and A. WARSHEL, *How important are entropic contributions to enzyme catalysis?*, Proc Natl Acad Sci U S A, 97 (2000), pp. 11899-904.

[490]   J. VILLA and A. WARSHEL, *Energetics and Dynamics of Enzymatic Reactions*, Journal of Physical Chemistry B, 105 (2001), pp. 7887-7907.

[491]   J. VINCZE, M. VALISKO and D. BODA, *The nonmonotonic concentration dependence of the mean activity coefficient of electrolytes is a result of a balance between solvation and ion-ion correlations*, J Chem Phys, 133 (2010), pp. 154507-6.

[492]   D. VOET and J. VOET, *Biochemistry*, John Wiley, Hoboken, NJ USA, 2004.

[493]   T. VORA, B. CORRY and S. H. CHUNG, *Brownian dynamics investigation into the conductance state of the MscS channel crystal structure*, Biochim Biophys Acta, 1758 (2006), pp. 730-7.

[494]   T. VORA, B. CORRY and S. H. CHUNG, *A model of sodium channels*, Biochim Biophys Acta, 1668 (2005), pp. 106-16.

[495]   L. VRBKA, M. LUND, I. KALCHER, J. DZUBIELLA, R. R. NETZ and W. KUNZ, *Ion-specific thermodynamics of multicomponent electrolytes: A hybrid HNC/MD approach*, J Chem Phys, 131 (2009), pp. 154109-12.

[496]   M. VROUENRAETS, J. WIERENGA, W. MEIJBERG and H. MIEDEMA, *Chemical modification of the bacterial porin OmpF: gain of selectivity by volume reduction*, Biophys J, 90 (2006), pp. 1202-11.

[497]   Y. WANG, L. XU, D. PASEK, D. GILLESPIE and G. MEISSNER, *Probing the Role of Negatively Charged Amino Acid Residues in Ion Permeation of Skeletal Muscle Ryanodine Receptor*, Biophysical Journal, 89 (2005), pp. 256-265.

[498]   A. WARSHEL, *Calculations of enzymatic reactions: calculations of pKa, proton transfer reactions, and general acid catalysis reactions in enzymes*, Biochemistry, 20 (1981), pp. 3167-77.

[499]   A. WARSHEL, *Computer simulations of enzyme catalysis: methods, progress, and insights*, Annu Rev Biophys Biomol Struct, 32 (2003), pp. 425-43.

[500]   A. WARSHEL, *Dynamics of enzymatic reactions*, Proc Natl Acad Sci U S A, 81 (1984), pp. 444-8.

[501]   A. WARSHEL, *Electrostatic origin of the catalytic power of enzymes and the role of preorganized active sites*, J Biol Chem, 273 (1998), pp. 27035-8.

[502]   A. WARSHEL, *Energetics of enzyme catalysis*, Proc Natl Acad Sci U S A, 75 (1978), pp. 5250-4.

[503]   A. WARSHEL, *Molecular dynamics simulations of biological reactions*, Acc Chem Res, 35 (2002), pp. 385-95.

[504]   A. WARSHEL and J. AQVIST, *Electrostatic energy and macromolecular function*, Annu Rev Biophys Biophys Chem, 20 (1991), pp. 267-98.

[505]   A. WARSHEL and M. LEVITT, *Theoretical studies of enzymatic reactions: dielectric, electrostatic and steric stabilization of the carbonium ion in the reaction of lysozyme*, J Mol Biol, 103 (1976), pp. 227-49.

[506]   A. WARSHEL, G. NARAY-SZABO, F. SUSSMAN and J. K. HWANG, *How do serine proteases really work?*, Biochemistry, 28 (1989), pp. 3629-37.

[507]   A. WARSHEL and A. PAPAZYAN, *Electrostatic effects in macromolecules: fundamental concepts and practical modeling*, Curr Opin Struct Biol, 8 (1998), pp. 211-7.

[508]   A. WARSHEL and W. W. PARSON, *Computer simulations of electron-transfer reactions in solution and in photosynthetic reaction centers*, Annu Rev Phys Chem, 42 (1991), pp. 279-309.

[509]   A. WARSHEL and S. T. RUSSELL, *Calculations of electrostatic interactions in biological systems and in solutions*, Quarterly Review of Biophysics, 17 (1984), pp. 283-422.

[510]   A. WARSHEL, S. T. RUSSELL and A. K. CHURG, *Macroscopic models for studies of electrostatic interactions in proteins: limitations and applicability*, Proc Natl Acad Sci U S A, 81 (1984), pp. 4785-9.

[511]   A. WARSHEL, P. K. SHARMA, Z. T. CHU and J. AQVIST, *Electrostatic contributions to binding of transition state analogues can be very different from the corresponding contributions to catalysis: phenolates binding to the oxyanion hole of ketosteroid isomerase*, Biochemistry, 46 (2007), pp. 1466-76.

[512]   A. WARSHEL, P. K. SHARMA, M. KATO, Y. XIANG, H. LIU and M. H. M. OLSSON, *Electrostatic Basis for Enzyme Catalysis*, Chemical Reviews, 106 (2006), pp. 3210-3235.

[513]   A. WARSHEL, F. SUSSMAN and J. K. HWANG, *Evaluation of catalytic free energies in genetically modified proteins*, J Mol Biol, 201 (1988), pp. 139-59.

[514]   T. WELTON, *Room-Temperature Ionic Liquids. Solvents for Synthesis and Catalysis*, Chem. Rev., 99 (1999), pp. 2071-2084.

[515]   S. T. WLODEK, J. ANTOSIEWICZ, J. A. MCCAMMON, T. P. STRAATSMA, M. K. GILSON, J. M. BRIGGS, C. HUMBLET and J. L. SUSSMAN, *Binding of tacrine and 6-chlorotacrine by acetylcholinesterase*, Biopolymers, 38 (1996), pp. 109-17.

[516]   M. R. WRIGHT, *An Introduction to Aqueous Electrolyte Solutions*, Wiley, New York, 2007.

[517]   L. XU, Y. WANG, D. GILLESPIE and G. MEISSNER, *Two Rings of Negative Charges in the Cytosolic Vestibule of Type-1 Ryanodine Receptor Modulate Ion Fluxes*, Biophysical Journal, 90 (2006), pp. 443-453.

[518]   Z. XU and W. CAI, *Fast Analytical Methods for Macroscopic Electrostatic Models in Biomolecular Simulations*, SIAM Review, 53 (2011), pp. 683-720.

[519]   H. YU, T. W. WHITFIELD, E. HARDER, G. LAMOUREUX, I. VOROBYOV, V. M. ANISIMOV, A. D. MACKERELL and B. T. ROUX, *Simulating Monovalent and Divalent Ions in Aqueous Solution Using a Drude Polarizable Force Field*, Journal of Chemical Theory and Computation, 6 (2010), pp. 774-786.

[520]   J. F. ZEMAITIS, JR., D. M. CLARK, M. RAFAL and N. C. SCRIVNER, *Handbook of Aqueous Electrolyte Thermodynamics*, Design Institute for Physical Property Data, American Institute of Chemical Engineers, New York, 1986.

[521]   C. ZHANG, S. RAUGEI, B. EISENBERG and P. CARLONI, *Molecular Dynamics in Physiological Solutions: Force Fields, Alkali Metal Ions, and Ionic Strength*, Journal of Chemical Theory and Computation, 6 (2010), pp. 2167-2175.

[522]   J. ZHANG, X. GONG, C. LIU, W. WEN and P. SHENG, *Electrorheological Fluid Dynamics*, Physical Review Letters, 101 (2008), pp. 194503.

[523]   Q. ZHENG, D. CHEN and G.-W. WEI, *Second-order Poisson Nernst-Planck solver for ion channel transport*, Journal of Computational Physics, 230 (2011), pp. 5239-5262.

[524]   Q. ZHENG and G.-W. WEI, *Poisson--Boltzmann--Nernst--Planck model*, J Chem Phys, 134 (2011), pp. 194101-17.

[525]   H. X. ZHOU and M. K. GILSON, *Theory of free energy and entropy in noncovalent binding*, Chem Rev, 109 (2009), pp. 4092-107.

[526]   S. ZHOU, Z. WANG and B. LI, *Mean-field description of ionic size effects with nonuniform ionic sizes: A numerical approach*, Physical Review E, 84 (2011), pp. 021901.

[527]   J. ZHU, E. ALEXOV and B. HONIG, *Comparative Study of Generalized Born Models: Born Radii and Peptide Folding*, The Journal of Physical Chemistry B, 109 (2005), pp. 3008-3022.

[528]   O. ZIKANOV, *Essential Computational Fluid Dynamics*, Wiley, New York, 2010.